\documentclass[prx,letterpaper,nobalancelastpage,twocolumn,superscriptaddress,nofootinbib,longbibliography]{revtex4-2}

\usepackage[utf8]{inputenc}
\usepackage{graphicx}
\graphicspath{ {./figures/} }
\usepackage{textcomp,gensymb}
\usepackage{lineno}

\usepackage[normalem]{ulem}
\usepackage{color}
\usepackage{units}
\usepackage{amsmath,amssymb}
\usepackage{comment}
\usepackage{hyperref}
\hypersetup{
    colorlinks=true,
    linkcolor=blue,
    filecolor=magenta,      
    urlcolor=cyan,
    citecolor=red,
    pdftitle={Hot Carrier Thermalization and Josephson Inductance Thermometry in a Graphene-based Microwave Circuit},
    pdfpagemode=FullScreen,
}

\definecolor{orange}{rgb}{1,0.5,0}

\renewcommand{\thesection}{\Roman{section}}
\newcommand{\caltechPH}{Department of Physics, California Institute of Technology, Pasadena, California 91125, USA}
\newcommand{\caltechAPH}{T. J. Watson Laboratory of Applied Physics, California Institute of Technology,
  1200 East California Boulevard, Pasadena, California 91125, USA}
\newcommand{\caltechIQIM}{Institute for Quantum Information and Matter, California Institute of Technology, Pasadena, California 91125, USA}

\begin{document}

\title{Hot Carrier Thermalization and Josephson Inductance Thermometry in a Graphene-based Microwave Circuit}

\author{Raj Katti}
\thanks{These authors contributed equally to this work.}
\affiliation{\caltechPH}
\author{Harpreet Arora}
\thanks{These authors contributed equally to this work.}
\affiliation{\caltechAPH}
\affiliation{\caltechIQIM}
\author{Olli-Pentti Saira}
\thanks{These authors contributed equally to this work.}
\affiliation{\caltechPH}
\affiliation{Brookhaven National Laboratory, Upton NY 11973, USA}
\author{Kenji Watanabe}
\affiliation{ National Institute for Materials Science, Namiki 1-1, Tsukuba, Ibaraki 305 0044, Japan}
\author{Takashi Taniguchi}
\affiliation{ National Institute for Materials Science, Namiki 1-1, Tsukuba, Ibaraki 305 0044, Japan}
\author{Keith C. Schwab}
\affiliation{\caltechPH}
\affiliation{\caltechAPH}
\author{Michael Roukes}
\affiliation{\caltechPH}
\affiliation{\caltechAPH}
\author{Stevan Nadj-Perge}
\email{Correspondence: s.nadj-perge@caltech.edu}
\affiliation{\caltechAPH}
\affiliation{\caltechIQIM}

\begin{abstract}

Due to its exceptional electronic and thermal properties, graphene is a key
material for bolometry, calorimetry, and photon detection. However, despite
graphene's relatively simple electronic structure, the physical processes
responsible for the transport of heat from the electrons to the lattice are
experimentally still elusive. Here, we measure the thermal response of
low-disorder graphene encapsulated in hexagonal boron nitride (hBN) by
integrating it within a multi-terminal superconducting device coupled to a
microwave resonator. This technique allows us to simultaneously apply Joule
heat power to the graphene flake while performing calibrated readout of the
electron temperature. We probe the thermalization rates
of both electrons and holes with high precision and observe a thermalization
scaling exponent consistent with cooling dominated by resonant electron-phonon
coupling processes occurring at the interface between graphene and
superconducting leads. The technique utilized here is applicable for wide range
of semiconducting-superconducting interface heterostructures and provides new insights 
into the thermalization pathways essential for the next-generation thermal detectors.
 
\end{abstract}

\maketitle

Graphene provides a tantalizing opportunity for the design and development of
bolometric detectors, due to its exceedingly small heat capacity
\cite{fongUltrasensitiveWideBandwidthThermal2012,
  fongMeasurementElectronicThermal2013}, much smaller compared to traditionally
synthesized thin films. In addition, the thermal conductivity of graphene can be
greatly changed by coupling it to superconducting or normal electrodes or
placing it on different substrates. Moreover, when graphene is contacted using
superconducting electrodes, the resulting Josephson coupling and the
corresponding supercurrents are highly dependent on electron temperature
\cite{borzenetsBallisticGrapheneJosephson2016}. Accordingly, graphene-based
Josephson junctions (gJJs) are particularly promising for detecting ultra-small
thermal responses at milli-Kelvin temperatures. In turn, gJJs can be tuned in
many ways, as graphene couples well with a variety of superconductors to form
highly transparent junctions, enabling supercurrents to persist over several
microns \cite{caladoBallisticJosephsonJunctions2015,
  draelosSupercurrentFlowMultiterminal2019}. Using different superconductors,
junction geometry, and operation at different carrier densities allows, in
principle, for a range of specific optimizations needed for detecting small heat
and optical signals. To achieve the highest sensitivity, for example, one can
choose to operate at the lowest temperatures and employ superconductors with a
small superconducting gap, similar to the approach that is taken in conventional
superconducting nanowire-based detectors. If a large dynamic range is required,
tuning the critical currents in graphene junctions by controlling carrier
density can provide additional flexibility in design.

\begin{figure*}[thp]
\begin{center}
    \includegraphics[width=0.9\textwidth]{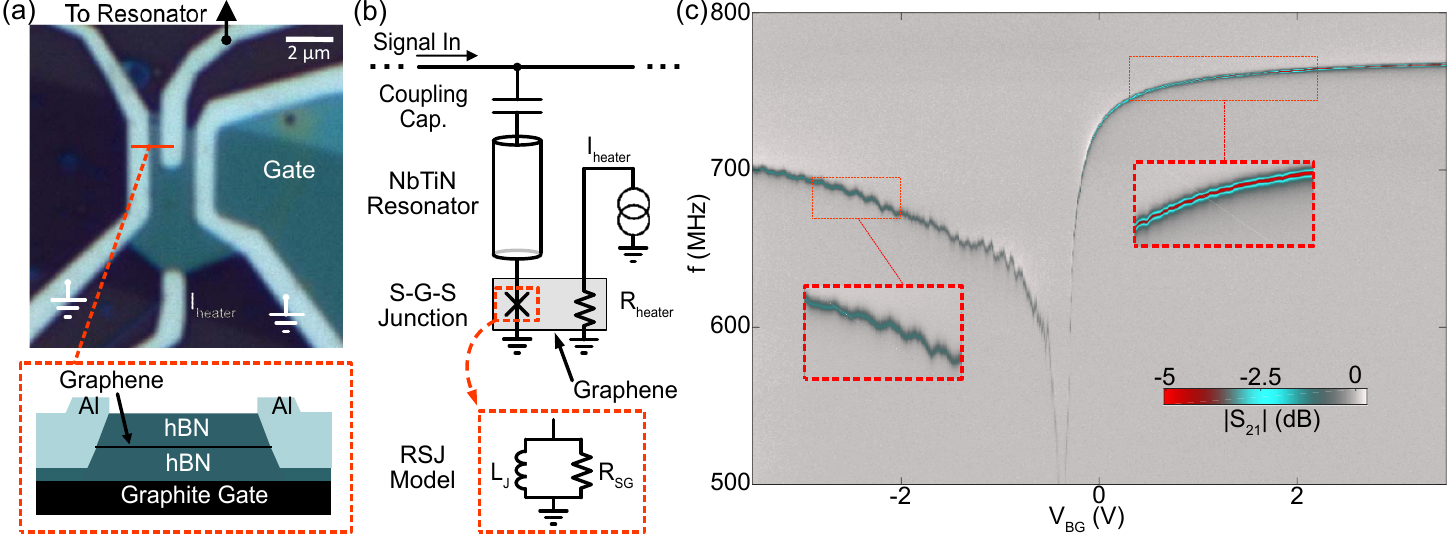}
\end{center}
\caption{Graphene Josephson junction and the characterization of the resonator
  circuit. (a), Optical image showing a top-down view of the graphene flake
  encapsulated in hexagonal boron nitride (blue-green) contacted by
  superconducting electrodes (light blue). The top contact is placed in close
  proximity to the ground wires to form the graphene Josephson junction (gJJ).
  The bottom contact placed far from the ground electrodes is used to apply
  Joule heating. The Inset shows the partial cross-section across the gJJ.
  Tuning the global carrier density in the graphene flake is achieved by
  applying a DC voltage $V_{BG}$ to a graphite backgate. (b), Simplified
  electrical circuit schematic (for full schematic see 
  Appendix~\ref{m:s21circuit}). A superconducting Niobium Titanium Nitride (NbTiN)
  resonator is coupled to the external microwave line via a coupling capacitor
  and terminated by the gJJ. The gJJ is electrically modeled as the parallel sum
  of a dissipationless branch of inductance $L_J=\frac{\Phi_0}{2\pi I_c}$ and a
  dissipative branch of resistance $R_{SG}$. A dedicated heater port allows
  application of Joule heat to the graphene flake. (c), $|S_{21}|$ vs. $V_{BG}$
  shows the evolution of the resonance feature. Near the charge neutrality point
  (CNP; $V_{CNP}= -0.3 \, \rm{ V}$), the gJJ maximally loads the resonator and,
  consequently, minimizes the resonant frequency. Far from the CNP, the gJJ acts
  as a low-impedance termination and maximizes the resonant frequency. On the
  hole-side ($V_{BG} < V_{CNP}$), Fabry-Perot type oscillations are visible due
  to formation of the regions of different doping in the bulk graphene (hole
  doping; p-type) and in vicinity of contacts (electron doping; n-type)
  \cite{schmidtBallisticGrapheneSuperconducting2018}.}

\label{fig: fig1}
\end{figure*}

Despite the significant progress in integrating graphene with superconducting
nanoelectronic devices, the present understanding of the thermalization of
electrons and holes in these systems is still incomplete. In most transport
measurements performed to date, thermalization in gJJs is thought to be
primarily driven by the electron-phonon interaction in graphene bulk
\cite{borzenetsPhononBottleneckGrapheneBased2013}, as the diffusion of unpaired
electrons into the metallic leads is suppressed due to the superconducting gap.
However, in the case where graphene is encapsulated within boron nitride (hBN),
deduced values of electron-phonon coupling from the experimental thermalization
rates \cite{leeGraphenebasedJosephsonJunction2020} are typically
orders-of-magnitude larger than theoretical predictions. Such a discrepancy is
not expected for materials with a simple band structure such as graphene, where
both the electronic and phonon spectrum can be readily calculated. Further,
recent scanning SQUID experiments, which provide spatially resolved thermal
imaging of graphene \cite{halbertalImagingResonantDissipation2017} have revealed
that, when electronic transport in graphene is ballistic, signatures of electron
thermalization are present only near physical edges, local defects, and close to
metallic contacts. However, such signatures of such boundary-mediated
thermalization have not so far been evident in transport measurements. Here we
present thermal measurements of a device architecture in which graphene
temperature is measured via changes in Josephson inductance
\cite{sairaDispersiveThermometryJosephson2016} caused by heating. In contrast to
typical critical current measurements that involve switching between
superconducting and resistive states, this approach allows to continuously
monitor thermal response with high precision that, in principle, depends only on
the measurement integration time. Surprisingly, for both electron and hole
doping, we observe temperature dependence of the thermal conductance consistent
with a resonant electronic scattering mechanism
\cite{kongResonantElectronlatticeCooling2018,
  tikhonovResonantSupercollisionsElectronphonon2018} that occurs at the
interface between graphene and superconducting leads.

Figure \ref{fig: fig1} shows a schematic of the device architecture and basic
characterization measurements. A gJJ is integrated into a graphene flake of
approximate area $\mathrm{A}=25 \, \mu\rm{m}^2$ (Fig.~\ref{fig: fig1}(a)). The
gJJ consists of a central superconducting contact separated from two
symmetrically placed superconducting contacts shorted to the ground plane.
Connection is made at the other end of the flake to a heater port used for
thermal characterization (see Appendix~\ref{m:fabrication} and
\ref{m:s21circuit} for details of device fabrication and the measurement
architecture). Superconducting aluminum is used for all contacts as it has a
small gap relative to other elemental superconductors; we expect this will
maximize temperature sensitivity in the sub-Kelvin temperature range of our
measurements. To probe the response of the gJJ supercurrent to changes in
electron density and temperature, we couple it to an on-chip resonator
\cite{schmidtBallisticGrapheneSuperconducting2018,
  wangCoherentControlHybrid2019} (Fig.~\ref{fig: fig1}(b)). Since the gJJ acts
as an additional inductive element, it modifies the resonant frequency, which we
monitor through microwave reflectometery. The parameters characterizing the gJJ,
the Josephson inductance $L_J=\frac{\Phi_0}{2\pi I_c}$ and subgap resistance
$R_{SG}$, depend strongly on electron density (see also Appendix~\ref{m:RSJmodel}). 
Accordingly, the resonant frequency and spectral width are
both highly dependent on the back gate voltage $V_{BG}$
\cite{schmidtBallisticGrapheneSuperconducting2018} (Fig.~\ref{fig: fig1}(c)).
Note that we can resolve the resonance over a large range of gate voltages; this
allows us to study phenomena arising from electron and hole doping as well as
near charge neutrality ($\mathrm{V_{BG}}\approx -0.4 \, \rm{V}$). For hole
doping ($\mathrm{V_{BG}} < -0.4 \, \rm{V}$), Fabry-Perot-type oscillations
indicate that carrier transport is ballistic in our high-quality graphene
sample.

\begin{figure}[thp]
\begin{center}
    \includegraphics[width=0.44\textwidth]{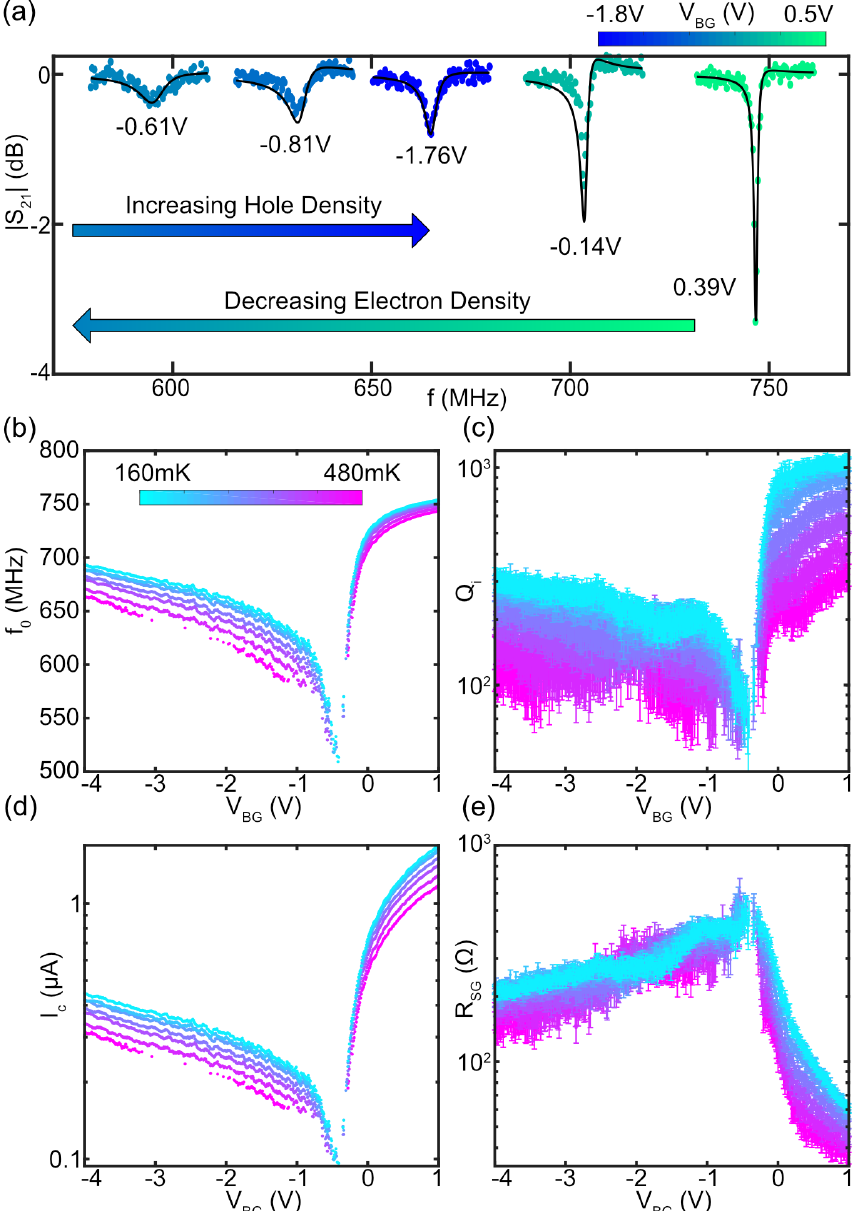}
\end{center}
\caption{ Resonance Fits and Extracted Parameters. (a), Representative
  $|S_{21}|$ data and fits for electron and hole doping. Color and labels denote
  backgate voltage $V_{BG}$. Arrows show the direction of resonant frequency
  shifts as $V_{BG}$ is swept from positive voltage (electron doped) to negative
  voltage (hole doped) through charge-neutrality. Green arrow shows the resonant
  frequency approaching $500 \,\rm{ MHz}$ as electron-doping decreases to charge
  neutrality. Blue arrow shows that the resonant frequency increasing away from
  $500 \,\rm{ MHz}$ as hole-doping increases. (b), $f_0$ as a function of
  $V_{BG}$ for representative $T_{mxc}$ = $160$, $210$, $260$, $310$, $360$,
  $410$, $460\, \rm{ mK}$. Extracted resonant frequency $f_0$ as a function of
  $V_{BG}$ shows a characteristic lineshape consistent with Fig.~\ref{fig:
    fig1}(c). As $T_{mxc}$ increases, $f_0$ decreases for all backgate voltages.
  (c), $Q_i$ as a function of $V_{BG}$ and $T_{mxc}$. (d, e), RSJ model
  parameters $I_c$ (d), and $R_{SG}$ (e) as a function of $V_{BG}$ and
  $T_{mxc}$. $I_c$ and $R_{SG}$ are determined using a numerical impedance model
  of the resonator/gJJ device with resonance parameters ($f_0$, $Q_i$) as inputs
  (see Fig.~\ref{exfig:rsj_model}).}
\label{fig: fig2}
\end{figure}

In addition to the electrostatic doping, the circuit resonance is also strongly
dependent upon temperature (Fig.~\ref{fig: fig2}). When the device temperature
increases, the resonance dip shifts to lower frequencies and broadens,
reflecting increased losses occurring within the junction. Importantly, the
observed shape of the resonance can be fitted using a standard four-parameter
Lorentzian fit function at all accessible carrier densities
($2.2\times10^{12} \, \mathrm{ holes/cm}^2<n_{carrier}<5.5\times10^{11}\, \mathrm{ electrons/cm}^2$)
and temperatures ($\mathrm{160 mK} < T_{mxc} < \mathrm{480 mK}$) (see also
Appendix~\ref{m:s21fitting}). The high level of agreement between data
and the fit (Fig.~\ref{fig: fig2}(a)) allows us to relate the deduced resonance
parameters to the physical properties of the junction. In particular, shifts of
resonant frequency $f_0$ and the overall resonance shape, which is set by
internal quality factor $Q_i$, can be related to parameters of the
resistively-shunted junction (RSJ) junction model
\cite{tinkhamIntroductionSuperconductivity2004}, the gJJ critical current $I_c$
and sub-gap resistance $R_{SG}$
\cite{schmidtBallisticGrapheneSuperconducting2018} (see Fig.~\ref{fig: fig1}(b)
and Appendix~\ref{m:RSJmodel}). These quantities determine the
small-signal electrical response of the junction at any temperature and doping
level. We note that an estimate of microwave losses in the junction is not
accessible from the switching current measurements that have typically been
employed in gJJ threshold detection schemes. Fitting the temperature dependence
of $I_c(T)$ allows estimation of an induced superconducting gap
$\Delta \sim 80 \, \rm{\mu eV}$ (see Appendix~\ref{m:gap}). Finally,
since we expect the resonator ringdown time $\tau$ to be the limiting time
constant in our device, we estimate from the fitted resonance parameters that
$\tau < 150 \, \rm{ns}$ for all backgate voltages (see
Fig.~\ref{exfig:rsj_model}).

\begin{figure*}[t!]
\begin{center}
    \includegraphics[width=0.9\textwidth]{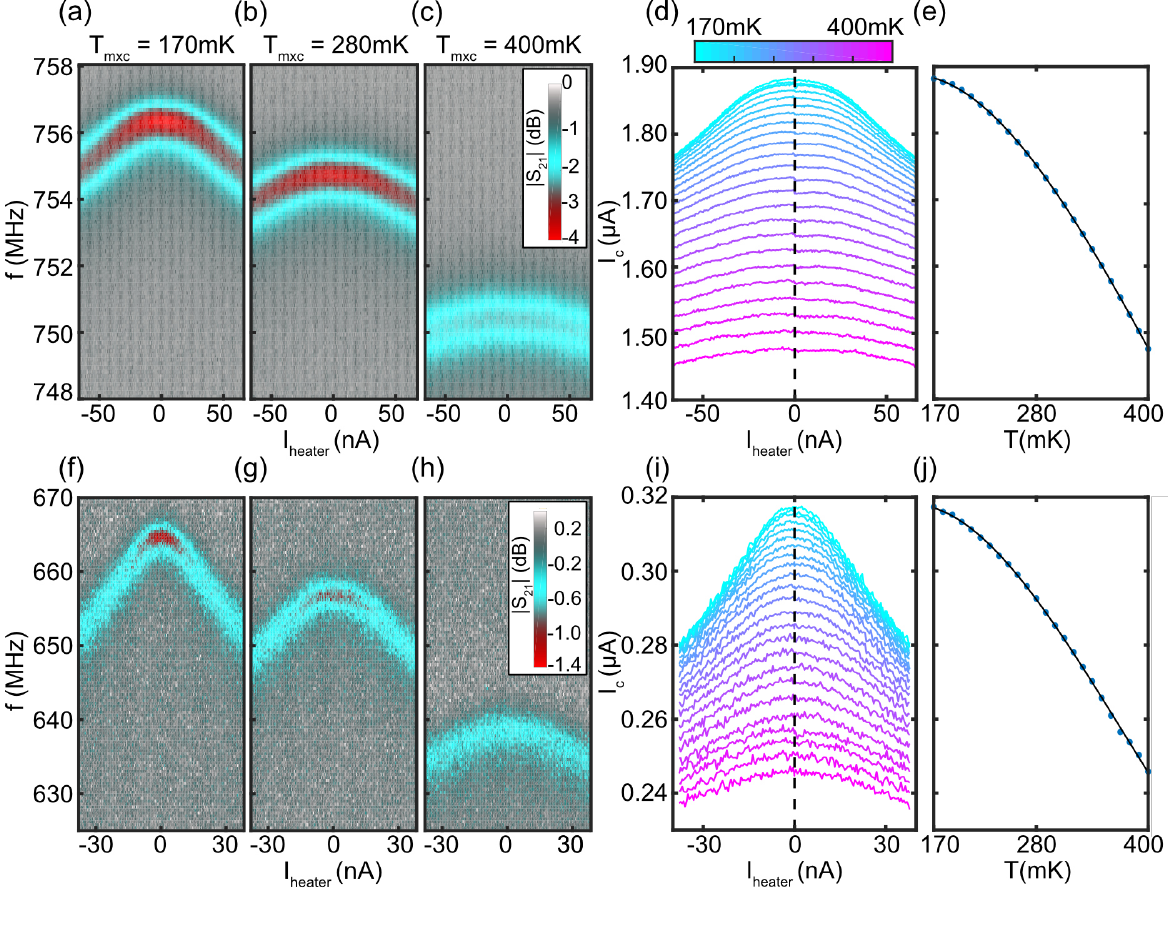}
\end{center}
\caption{ Electron-side and Hole-side Heating and Calibration. (a-c),
  Electron-side ($V_{BG}=1.1 \, \rm{ V}$) and (f-h) hole-side
  ($V_{BG}=-1.8 \, \rm{ V}$) $|S_{21}(f)|$ vs. $I_{heater}$ for three
  representative mixing stage temperatures (a, f) $T_{mxc}=170 \, \rm{mK}$ (b,
  g) $T_{mxc}=280 \, \rm{mK}$ (c, h) $T_{mxc}=400 \, \rm{mK}$. Applying a DC
  heater current $I_{heater}$ to the designated heater port decreases the
  resonant frequency of the device. As expected, the shifts are symmetric with
  respect to the polarity of $I_{heater}$. (d), Electron-side and (i) hole-side
  $I_c$ as a function of $I_{heater}$ for different mixing chamber temperatures.
  Fitting (a-c) and (f-h) allows extraction of resonance parameters ($f_0$,
  $Q_i$) and junction parameters ($I_c$, $R_{SG}$) (see also 
  Appendix~\ref{m:icrsg_fit}). The dashed line at $I_{heater}=0\, \rm{ nA}$ 
  corresponds
  to the data cut plotted in (e) and (j). (e) Electron-side and (j),
  hole-side calibration curve, the unheated $I_c$ as a function of $T_{mxc}$.
  Since $I_c$ monotonically decreases with increasing mixing chamber temperature
  $T_{mxc}$, there is a one-to-one correspondence between $I_c$ and graphene
  flake temperature.}
\label{fig: fig3}
\end{figure*}

To characterize the thermal properties of the gJJ device, we employ a
measurement configuration in which the gJJ is heated internally by applying a DC
current $I_{heater}$ to the heater port (Fig.~\ref{fig: fig3}). The port
electrode is placed sufficiently far from the ground electrodes to preclude
supercurrent flow. This configuration allows us to accurately monitor the input
power delivered to the graphene flake while simultaneously monitoring the
resonance frequency. For different device temperatures and doping,
representative changes in the $S_{21}$ resonance dip are shown in Fig.~\ref{fig:
  fig3}(a-c) and Fig.~\ref{fig: fig3}(f-h). By increasing the stage temperature
from $170 \, \rm{mK}$ to $400 \, \rm{mK}$, we observe a decrease in the resonant
frequency of $27 \, \rm{MHz}$ for holes, compared to $6 \, \rm{MHz}$ for
electrons. This is consistent with greater inductive loading (lower $I_c$) in
the hole regime (see Appendix~\ref{m:RSJmodel}). By applying a heater
current $I_{heater}$, the internal flake temperature $T$ is increased above
$T_{mxc}$, decreasing the resonant frequency. Combined with the measurements
taken at different temperatures for calibration (Fig.~\ref{fig: fig3}(e,j)) the
power vs. temperature characterization and, consequently, the thermal
conductivity $G_{th}$ of the graphene flake can be determined. We use this
approach to investigate thermal properties for both electron and hole doping
regimes.

\begin{figure*}[t!]
\begin{center}
    \includegraphics[width=\textwidth]{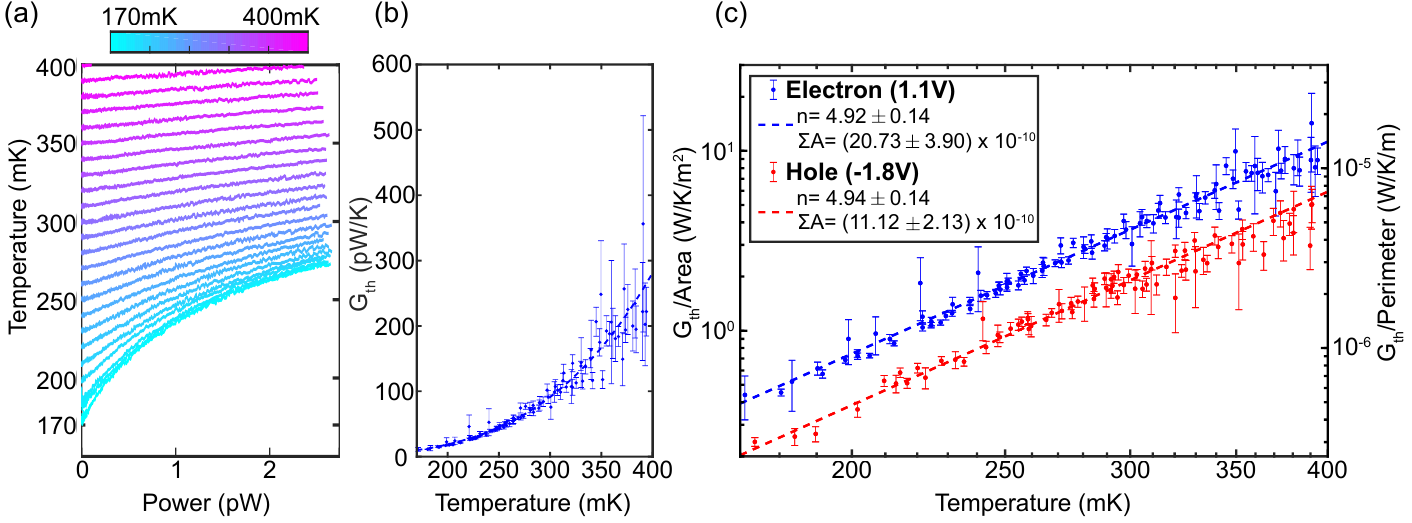}
\end{center}
\caption{Power-Temperature Curves. (a) Electron-side flake temperature as a
  function of heat power. From the injected DC current $I_{heater}$ and measured
  voltage drop $V$ across the heater port, the injected heater power can be
  determined $P_{heater} = I_{heater} \times V$. From the $I_c$ vs. $I_{heater}$
  traces in Fig.~\ref{fig: fig3}(d,i) and the $I_c$ vs. $T_{mxc}$ calibration in
  Fig.~\ref{fig: fig3}(e,j), flake temperature can be determined as a function
  of applied $P_{heater}$. Color corresponds to the mixing chamber stage
  temperature. (b) $G_{th}$ vs. $T_{mxc}$. Taking the numerical derivative
  $\frac{\partial P}{\partial T}$ of Fig.~\ref{fig: fig4}(a) allows the data to
  be plotted on a single line. Fit line is to the power law
  $G_{th} = \partial P/\partial T = n\Sigma AT^{n-1}$ where $n$ is the scaling
  exponent and $\Sigma A$ (in units of $\rm{W}/\rm{K^5}$) is the multiplicative
  factor. (c) Electron and hole $G_{th}$ vs. $T_{mxc}$ (log-log scale). Hole and
  electron doping both show power law scaling with an $n=5$ exponent.}
\label{fig: fig4}
\end{figure*}

The data we have acquired is consistent with a power law
$P_{heater} = \Sigma A(T^n - T_{mxc}^n)$, with electron temperature $T$, stage
temperature $T_{mxc}$, scaling exponent $n$ and the electron-phonon coupling
prefactor $\Sigma A$ (see also Appendix~\ref{m:ptfit}). We plot
$\partial P / \partial T = G_{th} = n \Sigma AT^{n-1}$ (Fig.~\ref{fig: fig4}(c))
which shows that the scaling exponents for hole and electron doping are
consistent with $n=5$. We note that our fitting procedures produce only
comparably small errors for each of the individual data points and, accordingly,
the uncertainty of the extracted scaling exponent is much less than $1$. This
enables us to clearly distinguish that the exponent obtained here is \emph{not}
consistent with the $n=3$ or $n=4$ scaling predicted for bulk electron-phonon
coupling in reduced dimensions \cite{viljasElectronphononHeatTransfer2010,
  chenElectronphononMediatedHeat2012}. While an $n=5$ scaling exponent is
expected for the electron-phonon coupling of a 3D electron gas
\cite{roukesHotElectronsEnergy1985}, these considerations do not apply for our
graphene device in which the electron and phonon density-of-states are 2D. Also, we 
note that the mechanism where hot electrons (or holes) diffuse into the superconducting 
aluminum leads before thermalization, while in principle possible, is not 
consistent with our observations (see Appendix~\ref{m:al_thermalization} 
for more detailed discussion). 

Measurements of hBN-encapsulated graphene performed previously
\cite{draelosSupercurrentFlowMultiterminal2019,
  leeGraphenebasedJosephsonJunction2020} reveal that $G_{th}$ (scaled by the
area) is about three orders-of-magnitude larger than predictions by simple bulk
electron-phonon coupling theory. The magnitude of
$G_{th}\sim 5-300 \, \rm{pW/K}$ in our measurements is consistent with these
observations. Due to enhanced mobility, hBN-encapsulated graphene is typically
in the ballistic scattering limit, in which the carrier mean free path $l_{mfp}$
is limited by the device dimension ($L_{device} \approx 5 \, \rm{\mu m}$ in our
sample). This observation has led to the hypothesis that the enhanced $G_{th}$
may arise from “resonant supercollisions”
\cite{kongResonantElectronlatticeCooling2018,
  tikhonovResonantSupercollisionsElectronphonon2018} a scenario consistent with
the spatially resolved measurements \cite{halbertalNanoscaleThermalImaging2016,
  halbertalImagingResonantDissipation2017}. In this scenario, defects located at
edge of the graphene flake locally enhance electron-phonon interactions and open
a thermalization pathway that dominates over electron-phonon coupling in the
bulk. Spatially-resolved scanning SQUID measurements show an enhancement of
surface phonon temperature at graphene edges and close to metal contacts. Theory
formulated to explain these results
\cite{tikhonovResonantSupercollisionsElectronphonon2018} suggests that an $n=5$
scaling exponent should hold down to milli-Kelvin temperatures ($T < T_{BG}$) in
the limit of strong scattering ($\delta \sim 1$). In this context, our high
precision measurements provide the first clear evidence that an $n=5$ scaling
exponent signaling that resonance supercollisions indeed dominate the
thermalization in graphene at sub-Kelvin temperatures. We note that further 
exploration of the device parameter space (e.g. sample size, aspect ratio, disorder) 
maybe needed to disentangle relations between different microscopic 
thermalization mechanisms in general.

We note that $G_{th}$ exhibits a power law consistent with $n=5$ for both electron
and hole doping, indicating that this mechanism remains dominant in both
regimes. Interestingly, the electron- and hole-side prefactors differ by a
factor of approximately two (see Fig.~\ref{fig: fig4}(c)). A possible
explanation for this difference arises from the energy distribution of resonant
scattering centers \cite{halbertalImagingResonantDissipation2017}. In this
scenario, different scattering centers are activated when the chemical potential
of the flake is shifted by the back-gate. Therefore, it is possible that the
difference in the prefactors can be attributed to different populations of
activated scatterers. Additionally, we note that, in the case of hole doping, the
intrinsic p-n junction formed between the graphene region close to the Al
contacts (which is always intrinsically n-doped) and the p-doped bulk may also
play a role. In this regime, holes from the bulk must pass across the p-n
junction in order to efficiently thermalize via resonant scattering centers.
Since the p-n junction has a finite transmission probability, it may therefore
reduce the overall thermalization rate. We note that attaining an accurate
calculation of the thermalization prefactor from first principles is difficult
due to effects outlined above and further theoretical work is needed for
quantitive comparisons.

In the context of detector technologies, graphene is argued to be a promising
platform for future scalable far-infrared or microwave detector-arrays
\cite{leeGraphenebasedJosephsonJunction2020,kokkoniemiBolometerOperatingThreshold2020}.
Its utility for this purpose is typically evaluated on the basis of optimization
of several key attributes including response time, responsivity, thermal
insulation and multiplexing that, in turn, require simultaneous optimization of
multiple device parameters. The hBN-encapsulated graphene devices studied here
provide large supercurrents and sub-microsecond response times that allow for
continuous monitoring of thermal response, and integration of the resonator
readout that permits straightforward frequency-division multiplexing of many
devices on a single feedline \cite{dayBroadbandSuperconductingDetector2003,
  wanduiThermalKineticInductance2020}. Moreover, in our scheme the presence of a
separate heater port can be employed for broad-spectrum energy detection. We
note that a thermal insulation of the architecture employed here can be achieved
at the expense of lowering the mobility in graphene by, for example, placing it
directly on the oxide substrate \cite{kokkoniemiBolometerOperatingThreshold2020}
instead of hBN.

Finally, we briefly compare the inductance readout scheme employed here with
graphene detectors based on junction switching
\cite{leeGraphenebasedJosephsonJunction2020,
  walshGrapheneBasedJosephsonJunctionSinglePhoton2017} (between the zero and
finite voltage state) as their potential applications may significantly differ.
The latter type of detectors register a “count” when the incident photon energy
is above a given threshold, and therefore forfeit the possibility of energy
spectroscopy provided by the linear, resonantly-coupled graphene detector
architecture pursued in this work. Further, threshold detectors intrinsically
provide slower response, which is limited by the cooling and resetting of the
junction after a photon absorption event. While this type of detector may be a
desirable option in the experiments where photon energy and arriving time is
known or controlled, the inductance readout detection scheme is more suitable
for novel spectroscopy applications of unknown sources
\cite{lara-avilaQuantumlimitedCoherentDetection2019}, including dark matter
detection \cite{hochbergDirectionalDetectionDark2017,
  kimDetectingKeVRangeSuperLight2020,
  mcallisterORGANExperimentAxion2017,baracchiniPTOLEMYProposalThermal2018} and
photon and phonon counting \cite{roukesYoctocalorimetryPhononCounting1999} where
linear response and ability to fully evaluate detection performance is important
(see Appendix~\ref{m:nep} for noise equivalent power characterization).

\noindent {\bf Acknowledgments:} We acknowledge useful discussions with Sophie
Li, Matt Matheney, Ewa Rej, and Jonas Zmuidzinas. {\bf Funding:} This work was
supported by NSF through program CAREER DMR-1753306 and Gist-Caltech memorandum
of understanding. S.N.-P. also acknowledges the support of DOE-QIS program
(DE-SC0019166), IQIM (NSF funded physics frontiers center) and the Sloan
foundation. M.L.R acknowledges support from NSF grant NSF-DMR-1806473.

\setcounter{section}{0}                                                                                                 
\renewcommand\appendixname{APPENDIX}           
\appendix                                           
\renewcommand\thesection{\Alph{section}}            
\renewcommand\thesubsection{\arabic{subsection}} 
\section{Fabrication}
\label{m:fabrication}

Fabrication of the superconducting resonator and coupling capacitor proceeds by
sputtering a few hundred nanometers of Niobium Titanitum Nitride (NbTiN) on an
undoped silicon wafer with 300 nanometers of thermal oxide. Typical
superconducting transition temperatures are $\sim$14 K. Subsequently, the
resonator and coupling capacitor are patterned by electron beam lithography
followed by an $\, \rm{SF_6}$ wet etch and Ar reactive ion etch. The graphene
heterostructure is assembled using standard exfoliation and stamping methods and
dropped on the resonator chip. 1-D edge contacts between the superconducting
metal and graphene heterostructure are patterned by electron beam lithography
followed by an Ar reactive ion etch and an electron beam evaporation of the
titanium adhesion layer and aluminum contacts.

\begin{figure}
\begin{center}
    \includegraphics[width=0.44\textwidth]{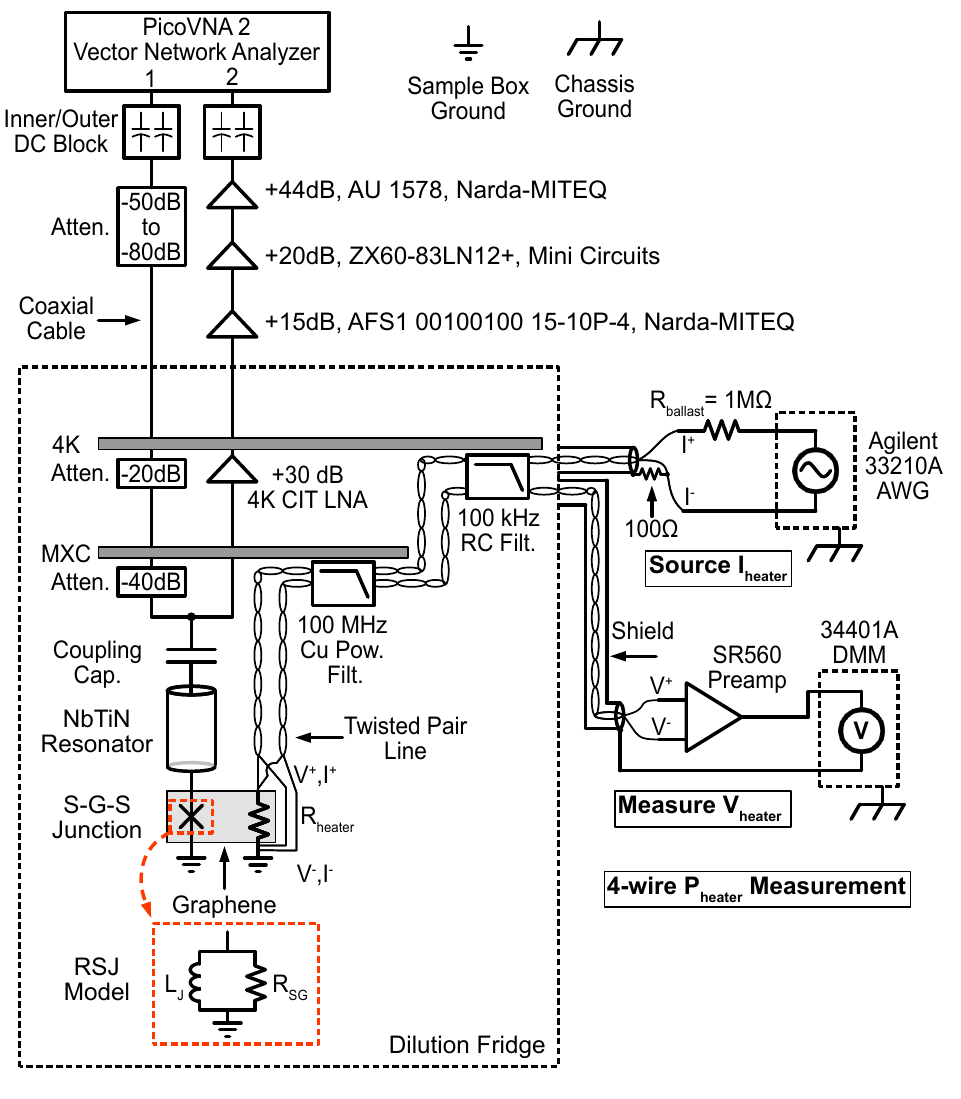}
\end{center}
\caption{$S_{21}$ Circuit Diagram. The circuit diagram shows both the
  $S_{21}$ readout of the resonance feature and the application and readout of
  $P_{heater}$. In the $S_{21}$ measurement, a signal is sourced from the
  PicoVNA2 vector network analyzer (VNA) and passes through a series of
  attenuators down to the resonator/graphene device held at $T_{mxc}$. The
  transmitted portion of the signal is amplified by an amplification chain with
  a first-stage 4K LNA and read out by the VNA. Application of $P_{heater}$
  proceeds by sourcing a current $I$ generated by a voltage sourced by the
  Agilent 33221A AWG and dropped over a $1 \, \rm{M}\Omega$ ballast resistor.
  After passing through a two stages of filters, the sourced current flows
  through the normal resistance $R_{heater}$ of the heater port of the graphene
  sample and dissipates Joule heat power in the flake. The voltage drop $V$
  across graphene heater is amplified by as SR560 preamp and read out by a
  32201A digital multimeter (DMM). In this way, the $P_{heater}=I \times V$
  delivered to the graphene flake is measured in a 4-wire measurement.}
\end{figure}

\section{$S_{21}$ Measurement Circuit}
\label{m:s21circuit}

A standard $S_{21}$ transmission measurement is performed in which a swept
microwave tone is sent out of Port 1 of a PicoVNA 2 vector network analyzer
(VNA) and down through attenuators and stages of the dilution refrigerator. The
impedance of the resonator/gJJ device loads the line and scatters the incoming
microwave tone. The transmitted portion of the microwave signal is amplified by
a first-stage $T_n=4\, \rm{K}$ CIT low noise amplifier, and then by three room
temperature amplifiers, where it is detected by Port 2 of the VNA.

To improve DC isolation between the device and the VNA, we include inner/outer
DC blocks on the ports of the VNA. To vary readout power incident upon to the
device, we vary the room temperature attenuation between $-50 \, \rm{dB}$ and
$-80 \, \rm{dB}$. The attenuation at the fridge stages ensures the
$300 \, \rm{K}$ noise at room temperature is attenuated below the noise floor of
the mixing chamber. In the diagram, the attenuators and amplifiers are
positioned immediately under the fridge stage to which they are thermally
anchored.

The heater measurements in Fig.~\ref{fig: fig3} and Fig.~\ref{fig: fig4} of the
main text are performed by applying a DC heater current $I_{heater}$ to the
heater port of the graphene flake and reading out the corresponding voltage drop
in a 4-wire measurement. To source $I_{heater}$, an Agilent 33210A AWG outputs a
DC voltage for the DC heating measurements of the main text and an AC voltage
for the noise equivalent power \cite{matherBolometerNoiseNonequilibrium1982}
measurements of section \ref{m:nep}. Since the ballast resistor
$R_{ballast} = 1 \, \rm{ M\Omega}$ is 3 orders-of-magnitude larger than the
heater port resistance $R_{heater} \approx 1 \, \rm{ k\Omega}$, the series
combination of the AWG and $R_{ballast}$ can be well-approximated as a current
source $I_{heater}$. The $I_{heater}$ current travels down PhBr twisted-pair
lines to the heater port where it Joule heats the graphene flake. Outside of the
fridge, the shield on the twisted pair lines is held at fridge ground. The
return line of the twisted pair is grounded through a $100 \, \Omega$ resistor
to a breakout box (not shown) which is also held at fridge ground. The return
line terminates at the negative terminal of the AWG. We note that the possible
ground loop introduced by the grounding of the twisted pair return line through
the $100 \, \Omega$ resistor does not have an appreciable effect on the
measurement.

\section{$S_{21}$ Fitting Procedure}
\label{m:s21fitting}

Fitting of the resonance feature follows the procedure in Ref.
\citenum{geerlingsImprovingQualityFactor2012}. Background-subtracted $S_{21}$
transmission data is fit to a four-parameter fitting function

\[
S_{21} = 1 - \frac{Q_0/Q_c - 2iQ_0\frac{\delta\omega}{\omega_0}}{1 + 2iQ_0\frac{\omega-\omega_0}{\omega_0}}
\]

Extracted fit parameters include resonant frequency $\omega_0$, internal quality
factor $Q_i$, coupling quality factor $Q_c$, and asymmetry parameter
$\delta\omega_0$. Total quality factor is defined as the parallel sum of the
dissipation channels $\frac{1}{Q_0} = \frac{1}{Q_i} + \frac{1}{Q_c} $. Error
bars in Fig. 2b-d correspond to the 95\% (2$\sigma$) confidence level calculated
from the covariance matrix of the fits. An asymmetry in the resonance circle can
cause the diameter of the resonance circle to occur off of the real axis. Such
an asymmetry may arise from a non-negligible line inductance or mismatched
input/output impedance.

\subsection{Resonance Dependence on $V_{BG}$}
\label{m:vbgtuning}

Figure \ref{fig: fig1}(c) shows how the resonance changes as a function of
$V_{BG}$. The maximal tuning of resonance frequency $f_0$ with $V_{BG}$ occurs
in the range $[V_{CNP}$, $V_{CNP} + 0.3 \, \rm{V}]$, where the
$\frac{\partial f_0}{\partial V_{BG}} \approx \frac{670 \, \rm{ MHz}}{1 V}$.
Assuming a parallel-plate capacitance of hBN ($\epsilon_{r}=3$) and a separation
$d=30 \, \rm{ nm}$ between the graphene flake and backgate,
$\frac{\partial f_0}{\partial n_{carrier}} \approx \frac{1.21 \, \rm{ GHz}}{10^{12}/\, \rm{cm}^2}$.
Since we estimate the area of our graphene flake to be $A=25 \mu m^2$, the
maximum sensitivity of our device used as an electrometer is
$\frac{\partial f_0}{\partial N_{carrier}} = \frac{4.84 \, \rm{ kHz}}{1\, \rm{e^-}}$.

\section{Fitting Procedure for Extraction of RSJ Parameters}
\label{m:icrsg_fit}
To deduce the physical parameters of the gJJ from the fit parameters of the
$S_{21}$ resonance feature, we employ an electrical impedance model of our
device which takes the inputs ($f_0$, $Q_i$) and numerically solves for junction
parameters ($I_c$, $R_{SG}$). $Sonnet \textsuperscript{\textregistered} 15.53$
is used to estimate the physical parameters of the NbTiN transmission line
resonator \cite{m.pozarMicrowaveEngineering4th2011} (See Table
\ref{table:icrsg_parameters}). The coupling capacitance $C_C$ is estimated by
fitting a set of resonances at $V_{BG}=-1.9\, \rm{V}$, numerically solving for
$C_C$, and creating a histogram of extracted $C_C$ values with mode
$C_c = 0.243 \, \, \rm{pF}$ and standard deviation of approximately
$\sigma_{C_c} = 0.02 \, \rm{pF}$.
\begin{figure}[htb]
\begin{center}
    \includegraphics[scale=1.5]{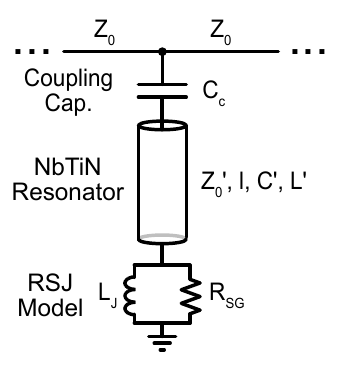}
\end{center}
\caption{Impedance Model. The electrical impedance model of the
  resonator-graphene device consists of the graphene Josephson junction in the
  RSJ model, a NbTiN transmission line resonator characterized by parameters in
  Table \ref{table:icrsg_parameters}, a coupling capacitor $C_c$, and 50
  $\Omega$ microwave ports.}
\label{exfig: icrsg_fit}
\end{figure}

\begin{table*}[htb]
\centering
\renewcommand*{\arraystretch}{1.75}
\begin{tabular}{ |l|l|c| }
\hline
 $C_C$ & Coupling capacitor & $0.243 \,\rm{ pF}$\\ 
 \hline
 $l$ & TLR length & $4989 \, \mu\rm{m}$\\ 
 \hline
  $C'$ & TLR capacitance per length& $3515 \,\rm{ pF/m}$\\ 
 \hline
  $L'$ & TLR inductance per length & $1130 \,\rm{ nH/m}$\\ 
 \hline
 $Z_0'$ & TLR characteristic impedance & $17.9 \, \Omega$\\ 
 \hline
  $v_{ph}$ & TLR phase velocity & $1.575\times 10^7 \, \rm{m/s}$\\ 
  \hline
  $Z_0$ & Reference characteristic impedance & $50 \, \Omega$\\ 
 \hline
   $Z_{out}$ & Parallel two-port impedance & $25 \, \Omega$\\ 
 \hline
\end{tabular}
\caption{Coupling Capacitor, Transmission Line Resonator (TLR), and Microwave Port Parameters.}
\label{table:icrsg_parameters}
\end{table*}

\section{Discussion of Extracted Parameters from Resonance Fits and RSJ Model}
\label{m:RSJmodel}

\begin{figure*}[t!]
\begin{center}
    \includegraphics[width=0.9\textwidth]{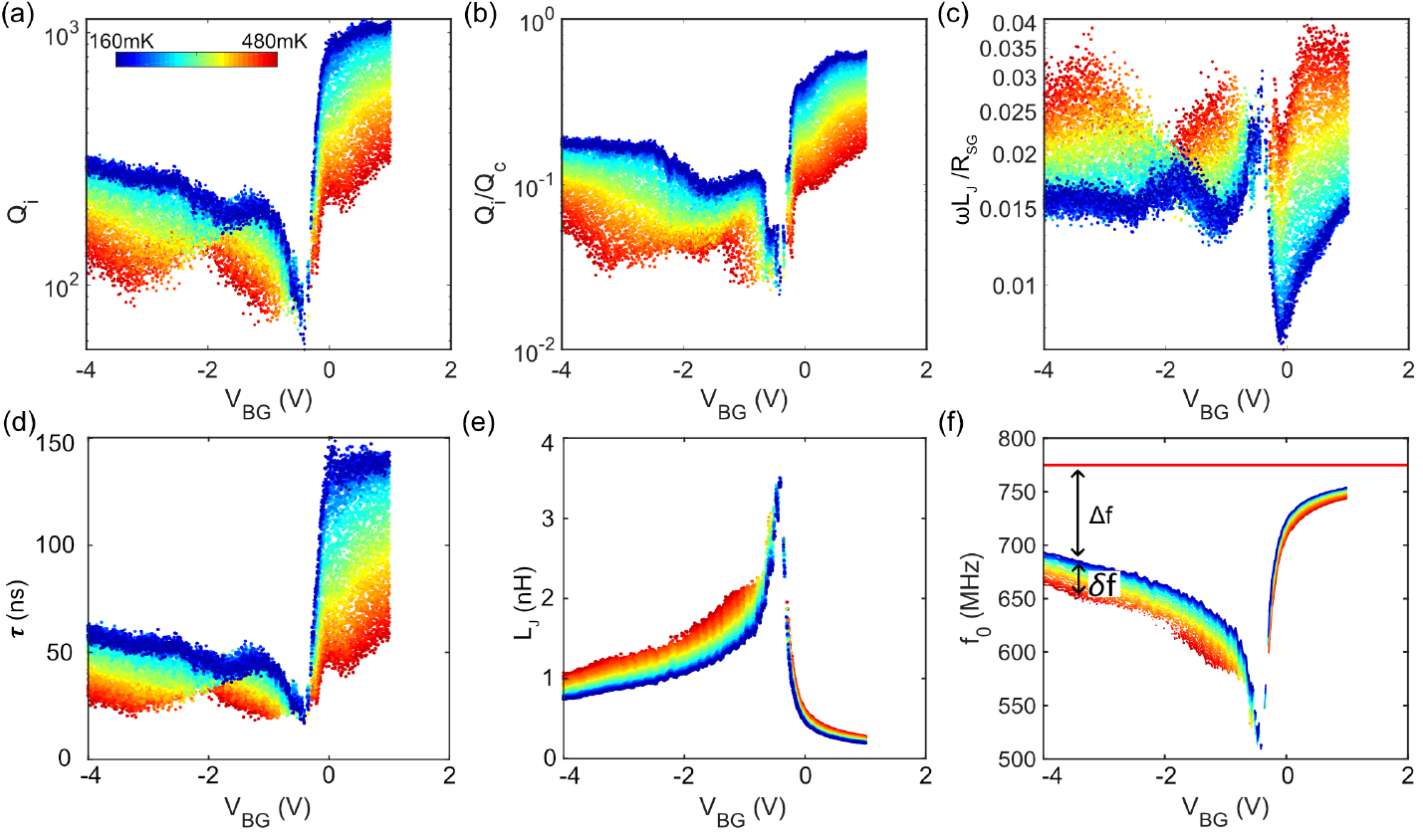}
\end{center}
\caption{Extracted Parameters from Resonance Fits and Impedance Model. (a) $Q_i$
  vs. $V_{BG}$. The internal quality factor $Q_i$ is extracted from the $S_{21}$
  fit function in \ref{m:s21fitting}. (b) $Q_i/Q_c$ vs. $V_{BG}$. Ratio of
  internal quality factor $Q_i$ and coupling quality factor $Q_c$ (also
  extracted from the $S_{21}$ fit function) shows that the device is in the
  undercoupled limit for all backgate voltages. (c) $\omega L_J/R_{SG}$ vs.
  $V_{BG}$. Ratio of the inductive branch impedance to resistive branch
  impedance in the RSJ model. (d) $\tau= Q_0/\omega_i$ vs. $V_{BG}$. The
  resonator time constant $\tau$ is expected to set the system time constant for
  all measured backgate voltages and temperatures. (e) $L_J$ vs. $V_{BG}$. The
  Josephson inductance $L_J = \frac{\Phi_0}{2\pi I_c}$. (f) $f_0$ vs. $V_{BG}$.
  The red line corresponds to the projected unloaded ($L_J=0 \, \rm{nH}$)
  resonance frequency. $\Delta f$ corresponds to the loaded
  ($L_J \neq 0 \, \rm{nH}$) resonance frequency at $T_{mxc}=160\, \rm{mK}$.
  $\delta f$ corresponds to further shift in the resonance frequency due to the
  increase in flake temperature.}
\label{exfig:rsj_model}
\end{figure*}

As shown in Fig.~\ref{exfig:rsj_model}, our fitting and modeling procedure
allows several fit and junction parameters to be plotted as a function of
backgate voltage $V_{BG}$ and flake temperature $T_{mxc}$.

Figure~\ref{exfig:rsj_model}(a) shows a dip in $Q_i$ at $V_{BG}=-2\rm{V}$, which
is propagated to the other plots Fig.~\ref{exfig:rsj_model}(b-d). This dip
arises from an asymmetry in the $S_{21}$ parameter which rotates the resonance
circle off the real axis. Such rotations can arise from line impedance
mismatches and parasitic couplings \cite{geerlingsImprovingQualityFactor2012}.
Since $R_{SG}$ is determined primarily by $Q_i$, $R_{SG}$ is sensitive to
dissipation in the graphene flake as well as the electromagnetic environment of
the flake/resonator assembly. By contrast, $f_0$ and $I_c$ are largely
insensitive to these effects, so our thermometry based upon the dispersive
shifts of the resonance is also largely insensitive to these effects.

Figure~\ref{exfig:rsj_model}(b) shows that our device for all backgate voltages
is in the undercoupled limit ($Q_i < Q_c$), where dissipation occurs primarily
within device instead of via the coupling to the microwave lines. The variation
of the coupling quality factor $Q_c$ is consistent with the circuit model and a
constant coupling capacitor $C_c= 0.243$ pF.

The dispersive shifts of the resonance can be understood from the impedance
model shown in Fig.~\ref{exfig: icrsg_fit}, which consists of a transmission
line resonator terminated by the junction impedance. This model predicts an
unloaded ($L_J = 0 \, \rm{ nH}$) resonant frequency of
$f_{unload}=774.75 \, \rm{ MHz}$ as indicated by the solid red line in
Fig.~\ref{exfig:rsj_model}(f). When a finite inductance $L_J$ loads the
transmission line resonator, the resonant frequency decreases. This occurs
because a change in the terminating impedance alters the boundary condition at
the terminating end of the resonator. In the case of the unloaded resonator,
i.e. a $\lambda/4$ resonator, the termination is a short-to-ground, which fixes
the boundary voltage at $V=0$. This enforces the resonance condition that the
length of the resonator equals one quarter of the resonant wavelength, i.e.
$\lambda/4 = l$. However, terminating the transmission line resonator in an
inductance alters the boundary condition such that the boundary voltage
amplitude is fixed at some $V=V_0>0$. This has the effect of enforcing the
resonance condition that a quarter-wavelength is larger than the resonator
length, i.e. $\lambda/4 > l$, or, analogously, that the resonant frequency is
decreased relative to the unloaded case. The larger the terminating impedance,
i.e. the larger $L_J$, the lower the resonant frequency
\cite{schmidtBallisticGrapheneSuperconducting2018,
  m.pozarMicrowaveEngineering4th2011}.

Due to higher contact transparency, electron doping should exhibit a larger
supercurrent than hole doping. It follows that the electron side should exhibit
a smaller $L_J$ than the hole side, and, correspondingly, the electron side
should exhibit a smaller decrease in resonant frequency relative to $f_{unload}$
than the hole side. This is consistent with Fig.~\ref{exfig:rsj_model}(f) for
electron and hole doping, i.e. $\Delta f_{electron} < \Delta f_{hole}$ where
$\Delta f$ is defined as the resonant frequency decrease at
$T_{mxc}=160 \, \rm{mK}$.

Increasing the flake temperature further increases $L_J$ and decreases the
resonant frequency. A rough estimate of the further decrease of the resonant
frequency $\delta f$ due to increased temperature is as follows:

	\[\frac{|\delta f|}{|\Delta f|}  \approx \frac{|\delta I_c |}{|\Delta I_c|} \] 

  As discussed in the section \ref{m:gap}, $I_c$ typically decreases by 20-30\%
  as the flake temperature is increased from $160 \, \rm{mK}$ to
  $400 \, \rm{mK}$. From the main text Fig.~\ref{fig: fig3},

Hole Side: 
\[\frac{|\delta f_{hole}|}{|\Delta f_{hole}|} = \frac{26 \, \rm{ MHz}}{110 \, \rm{ MHz}} \approx 24\%\]

Electron Side:
\[\frac{|\delta f_{electron}|}{|\Delta f_{electron}|} = \frac{5.9\, \rm{ MHz}}{18.6 \, \rm{ MHz}} \approx 32\%\]

The change in resonant frequency is therefore consistent with the typical change
in $I_c(T)$. We conclude that the greater magnitude of frequency decrease on the
hole side relative to the electron side follows as a straightforward result of
the greater inductive loading of the transmission line resonator.

As shown in Fig.~\ref{exfig:rsj_model}(c), $\frac{\omega_0L_J}{R_{SG}}$ is a
common figure-of-merit for RF-driven Josephson junctions
\cite{vanduzerPrinciplesSuperconductiveDevices1998}. It compares the impedance
of the dissipationless supercurrent branch to the dissipative resistive branch.
A smaller value of $\frac{\omega_0L_J}{R_{SG}}$ denotes a less dissipative
device. At $T_{mxc}=160 \, \rm{mK}$, $\frac{\omega_0L_J}{R_{SG}} \approx 1.5\%$
within a factor of 2. As the temperature rises to $T_{mxc}=400\,\rm{ mK}$,
$\frac{\omega_0L_J}{R_{SG}}$ increases to 3\%. This is consistent with decreases
in $I_c$ raising the impedance of the dissipationless branch and driving more
current through the dissipative branch, as indicated by the degrading quality
factor with increasing flake temperature (see Fig.~\ref{fig: fig2}(c)).

\section{$I_c$ vs. $T_{mxc}$ Fits and Extraction of Induced Superconducting Gap}
\label{m:gap}

\begin{figure*}
\begin{center}
    \includegraphics[width=0.9\textwidth]{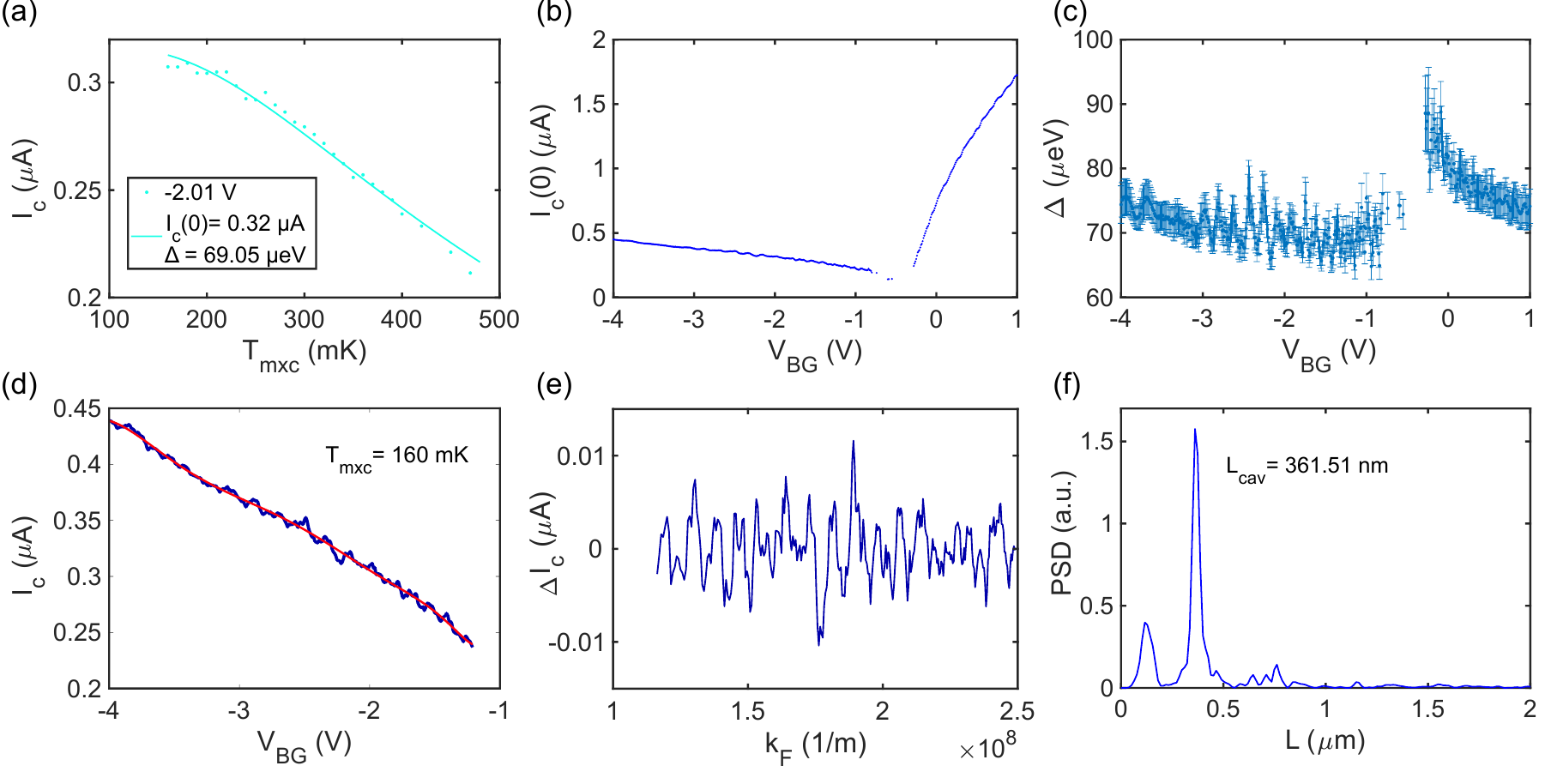}
\end{center}
\caption{(a) $I_c$ vs. $T$. An example fit of $I_c$ vs. $T$ for
  $V_{BG}= -2.01 \, \rm{V}$ with extracted fit parameters $I_c(0)$ and $\Delta$.
  (b) Fit parameter $I_c(0)$ vs. $V_{BG}$. $I_c(0)$ fit parameter is shown for
  both electron and hole doping. (c) Fit parameter $\Delta$ vs. $V_{BG}$. A
  coarse estimate of induced gap $\Delta \approx 80 \, \rm{\mu V}$. Fine
  features are discussed in the text. (d) Hole side $I_c$ vs. $V_{BG}$. Blue
  trace is hole side $I_c$ data for $T_{mxc}=160 \, \rm{mK}$. Red trace is the
  slowly-varying background as fit to a $7^{\rm{th}}$-order polynomial. (e)
  Background-subtracted $\Delta I_c$ vs. $k_F$. $\Delta I_c$ is obtained by
  subtracting the two traces in Fig.~\ref{exfig: gap}(d). (f) Power spectral
  density of $\Delta I_c$. The large peak is consistent with an effective
  Fabry-Perot cavity length of $L_{cav}=361.51 \, \rm{nm}$. }
\label{exfig: gap}
\end{figure*}

Due to the measurement architecture employed here, we cannot perform 4-wire
measurements directly on the gJJ to estimate the induced superconducting gap
$\Delta_0$. Instead, we perform a fitting procedure based upon the temperature
dependence of the critical current $I_c(T)$.

The $I_c(T)$ vs. $V_{BG}$ data in Fig.~\ref{fig: fig2}(d) is fit to extract
physical parameters. The fit function we employ describes the supercurrent that
arises from thermally populating the Andreev bound states (ABS) in a ballistic
junction
\cite{leeUltimatelyShortBallistic2015}. \[ I_c(T) = I_{c}(0) \tanh\Big(\frac{\Delta}{2 k_B T}\Big) \]
The two fit parameters correspond to the physical parameters $I_{c}(0)$, the
zero-temperature critical current, and $\Delta$, the induced superconducting
gap. An example fit is shown in Fig.~\ref{exfig: gap}(a).

In Fig.~\ref{exfig: gap}(b), modulation of the fit parameter $I_c(0)$ with
$V_{BG}$ on the hole side is consistent with $pnp$-type Fabry-Perot interference
as discussed in the main text and Fig.~\ref{fig: fig1}(c). Following the
standard method for determining Fabry-Perot cavity length in ballistic graphene,
we subtract the slowly varying background with a fit to a $7^{\rm{th}}$-order
polynomial (see Fig.~\ref{exfig: gap}(d-e)) and take the power spectral density
(see Fig.~\ref{exfig: gap}(f)). The large peak in the power spectral density is
consistent with a Fabry-Perot cavity length of $L_{cav}=361.51 \, \rm{nm}$.
Structure on the electron side could be caused by an $nn'n$-type Fabry-Perot
cavity \cite{nandaCurrentPhaseRelationBallistic2017}.

From Fig.~\ref{exfig: gap}(c), we can make a coarse estimate of the induced
superconducting gap $\Delta \approx 80 \, \rm{\mu V}$. However, further
measurements are needed to determine whether the finer structure of
Fig.~\ref{exfig: gap}(c) is due to the physics of the S-G-S junction or an
artifact of the fitting procedure. Toward this end, it would be useful to
perform simultaneous RF characterization and DC multiple-Andreev reflection
measurements on a gJJ sample \cite{schmidtBallisticGrapheneSuperconducting2018}.

\section{Power vs. Temperature Fitting Procedure}
\label{m:ptfit}
To obtain the $G_{th} = \frac{\partial P}{\partial T}$ vs. $T$ in Fig.~\ref{fig:
  fig4}(c), we first perform piece-wise linear fits of the $P-T$ curves of
Fig.~\ref{fig: fig4}(a). Subsequently, we perform a nonlinear least squares fit
of the $G_{th}$ vs. $T$ to the fitting function

\begin{equation}
    G_{th} = n \Sigma A T^{n-1}
\end{equation}

with free fit parameters $n$ the scaling exponent and $\Sigma A$ the
electron-phonon coupling. The errors in the free fit parameters correspond to
the $2\sigma$ (95\%) errors obtained from the nonlinear least squares fit. On
the electron side, we include an exclusion criteria at the limit of the
temperature resolution of our device. This criteria does not appreciably change
the extracted $n$ or $\Sigma A$. Without the exclusion criteria the extracted
fit parameters are $n=5.04 \pm 0.2$ and
$\Sigma A=(25.25 \pm 6.89)\times 10^{-10} \, \rm{ W/K}^5$. With the exclusion
criteria, the extracted fit parameters are $n=4.92 \pm 0.14$ and
$\Sigma A=(20.73 \pm 3.90)\times 10^{-10} \, \rm{ W/K}^5$.

\begin{figure*}
\begin{center}
    \includegraphics[width=0.9\textwidth]{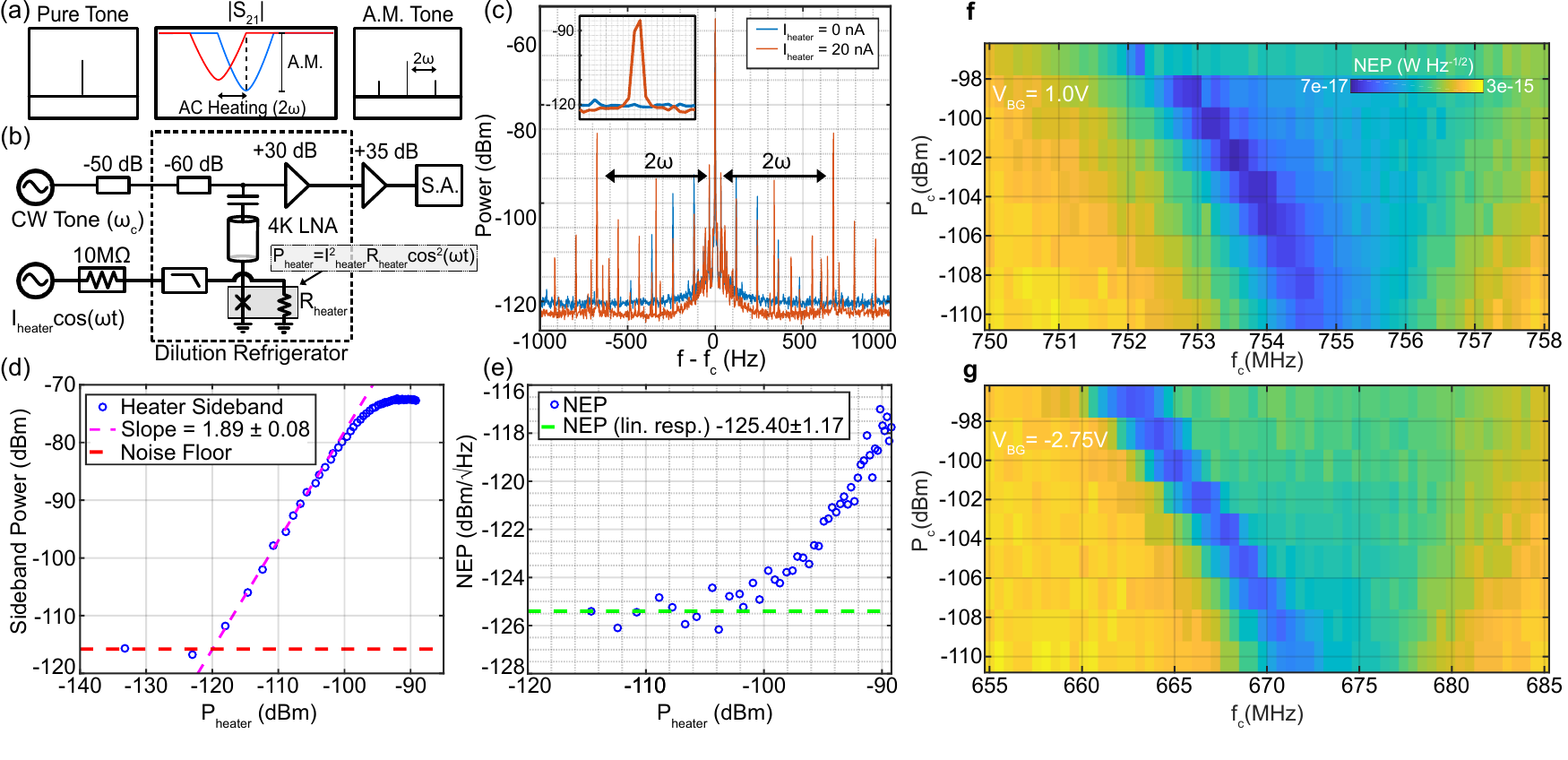}
\end{center}
\caption{Noise Equivalent Power (NEP) characterization. (a) Schematic of
  Measurement Chain. First panel shows a pure carrier tone sent down the
  microwave line. Second panel shows that an applied $\omega$ heater current and
  subsequent $2\omega$ modulation of the heat power (and temperature of the
  graphene flake) yields a $2\omega$ modulation of the transmission function
  ($S_{21}$ parameter) between unheated (blue) and heated (red) states. The pure
  tone (dashed line) is placed within the bandwidth of the transmission function
  and amplitude modulated at $2\omega$ with a modulation index that depends on
  the magnitude of the $S_{21}$ dip. Third panel shows the amplitude-modulated
  signal with sidebands at $2\omega$ as it appears on the spectrum analyzer. The
  measured signal-to-noise ratio of the sideband is used to determine the $NEP$.
  (b) Circuit Diagram. A continuous-wave carrier tone at $\omega_c$ is sent down
  a microwave line to the graphene device, amplified, and read out by a spectrum
  analyzer. An AC heater current at frequency
  $\omega = 2\pi \times 337 \, \rm{Hz}$ injects a $2\omega$ heat power
  $P_{heater}$ in the graphene flake and produces $2\omega$ amplitude modulation
  of the carrier tone, as discussed in (a). (c) Representative spectrum at
  output of measurement chain. Spectrum as read out by spectrum analyzer
  ($\rm{RBW}= 1 \, \rm{Hz}$) for applied heat power off (blue) and on (red). The
  primary effect of the applied heat is to produce sidebands spaced at
  $2 \omega$ from the the carrier tone. Other peaks in the spectrum exist at
  multiples of the line frequency. A peak at $\omega$ is consistent with a DC
  offset in the applied heat power. Inset shows the $2\omega$ sideband. (d)
  Sideband Power vs. $P_{heater}$. In the low-$P_{heater}$ linear-response
  regime, the sideband voltage $V_{sb} \propto P_{heater}$. Since the spectrum
  analyzer reads out the sideband power, $P_{sb} \propto P_{heater}^2$, which is
  consistent with the slope at low $P_{heater}$. (e) $NEP$ vs. $P_{heater}$. The
  linear-response regime is characterized by a regime of constant $NEP$, before
  rising as the amplitude modulation saturates to its maximal value. The $NEP$
  plotted in (g,f) corresponds to the linear response regime (green dashed
  line). (f, g) $NEP$ vs. carrier power $P_c$ and carrier frequency $f_c$ for
  (f) electron-side ($V_{BG}=1.0\, \rm{V}$) and (g) hole-side
  ($V_{BG}=-2.75\, \rm{V}$). Minimal $NEP$ occurs near the resonance dip minimum
  where amplitude modulation is largest. As carrier power $P_c$ is increased,
  the resonance dip downshifts to lower frequencies and is driven into
  nonlinearity, as characterized by an asymmetric resonance lineshape with steep
  falling edge and shallow rising edge. The minimum $NEP$ tracks the steep
  falling edge where amplitude modulation is greatest.}
\label{exfig: nep}
\end{figure*}

\section{Other thermalization pathways}
\label{m:al_thermalization}

In this section, we briefly discuss alternative 
thermalization pathways that can occur in our experimental geometry. 
While they indeed occur, we note that the thermal conductance corresponding 
to these alternative pathways 
are all too small to explain our measurements. 

\noindent {\textbf{Thermalization via bulk phonons}}

\begin{figure*}
    \includegraphics{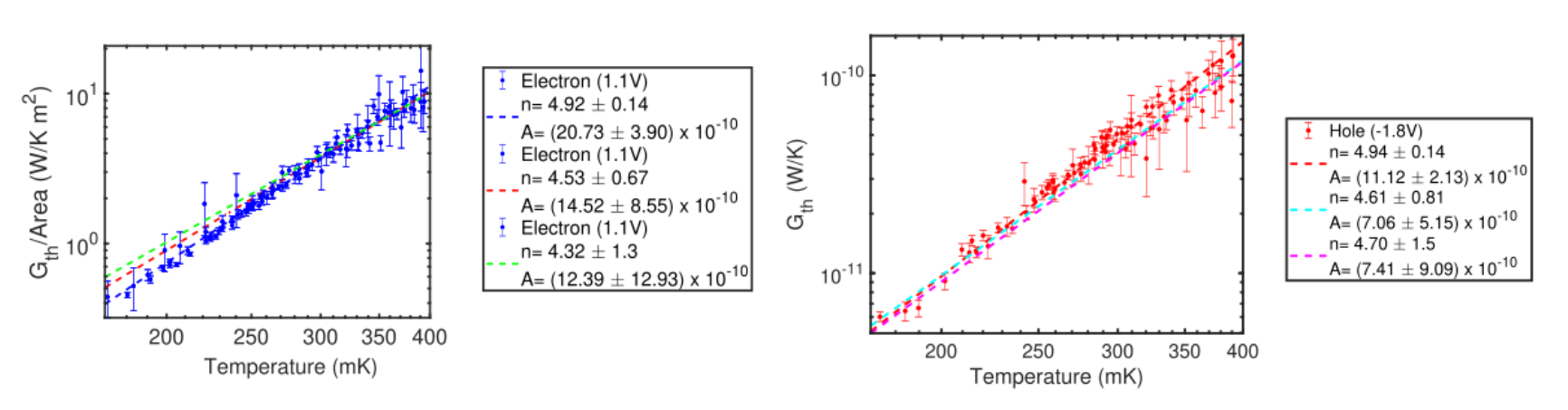}
\caption{The fits for the electron and hole doping in different temperature ranges. Electron
doping: Blue dashed line is the original full fit from 180 mK to 400 mK. Red is 300 mK to 400mK. Green is 330 mK
to 400 mK. Hole doping: Red dashed line is the full data fit. Cyan is 300 mK-390 mK. Magenta is 330 mK-390 mK.
For both electron and hole doping the n decreases (but stays well above n>4) when part of the data is used for
fitting. This indicates that $n = 4$or $n = 3$ power exponents are inadequate to describe our data. }
\label{exfig: fit_ranges}
\end{figure*}

The bulk phonons are often invoked as the main source of electron thermalization 
in graphene. However, besides having a different exponent ($n=3$ or $n=4$ not agreeing 
with our data, see Fig. \ref{exfig: fit_ranges}), the cooling 
rate via bulk graphene phonons is too small to explain experimental findings. 
As discussed in previous literature, the typical thermal conductance expected from bulk phonons
is two orders of magnitude smaller than the measured data. We note that, in this context of 
the overall cooling rate, our measurements are roughly in line with previous graphene-hBN experiments. 

\begin{figure*}
    \includegraphics{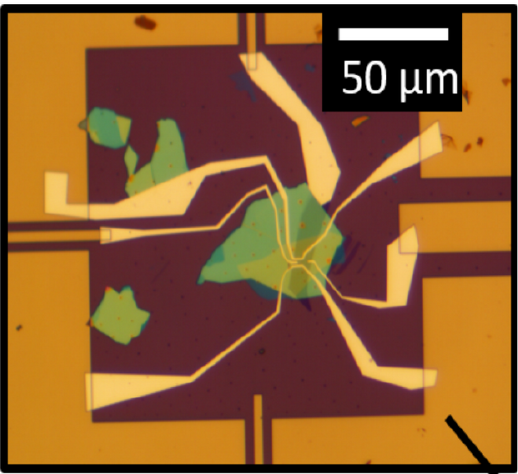}
\caption{The optical image showing the
area covered in Aluminum and NbTiN. Considering that Al thickness is
100nm the total volume of Aluminum part is $V= 1.5\times 10^{-16} $~m$^3$ approximately. This small Al volume limits
the amount of thermalization that can be achieved through contacts. }
\label{exfig: device}
\end{figure*}

\noindent {\textbf{Thermalization in Aluminum leads}:} Another possibility is that 
the hot electrons enter Al leads. While the tunneling of electrons (or holes) in a 
superconductor is expected to be suppressed due to the existence of a finite single-particle gap, 
previous work found that this process can still be sizable when using Aluminum electrodes. 
The critical distinction between our study and previous work using Aluminum contacts (see, 
for example, Ref. \cite{voutilainenEnergyRelaxationGraphene2011}), is that we used Al only 
as immediate contact to graphene (Fig. \ref{exfig: device}), and beyond that, the electrodes used in our experiment 
are made from Niobium Titanium Nitride (NbTiN) which has a much larger superconducting 
gap ($\Delta\approx1$~meV). In this context, the amount of heat that could “leak” into 
Aluminum contacts is much lower compared to measured values of thermal conductance. 
A total volume of 
Aluminum in our device is only approximately $V= 150 \times 10^{-18}$~m$^3$. By considering the established value for electron-phonon coupling in Aluminum ($\Sigma = 0.3\times10^9$~W/K$^5$m$^3$, see Ref. \cite{giazottoOpportunitiesMesoscopicsThermometry2006}) the corresponding P. vs T
dependence is expected to follow $P \approx 0.98 \times \Sigma \times V \times T^5 \times exp(-\Delta/(kT))$. Here $\Delta \approx 170-200$ $\micro$eV is Aluminum
superconducting gap measured in our experiments and a numerical pre-factor of 0.98 is estimated in Ref. \cite{timofeevRecombinationLimitedEnergyRelaxation2009}. At $T = 200$~mK this rate corresponds to
$G=dP/dT=0.25\times10^{-12}$~W, approximately two orders of magnitude smaller than the observed thermalization 
rates at our lowest temperatures. 

\section{Noise Equivalent Power}
\label{m:nep}
\subsection{Theory}

A key figure-of-merit for linear power detectors is noise-equivalent power
($NEP$). A power-to-voltage detector has a responsivity $\mathcal{R}$, such that
	\[V_{out} = \mathcal{R}(P_{in})\]

  In the linear-response regime, i.e. for small applied power, this expression
  simplifies to

\[ \delta V_{out} \approx \bigg(\frac{\partial V_{out}}{\partial P_{in} } \bigg|_{\delta P_{in}=0} \bigg) * \delta P_{in} \]

In this regime, the $NEP$ of a power-to-voltage detector (in units of
$\frac{W}{\sqrt{Hz}}$) can be defined as that power spectral density at the
device input which produces the measured voltage spectral density $\sqrt{S_V}$
at the output:

\[  NEP  \equiv \frac{\sqrt{S_V}} {\partial V_{out}/\partial P_{heater}\rvert_{\delta P_{heater}=0} }\]

The above expression suggests two immediate ways to measure the $NEP$. One is to
measure the voltage spectral density $\sqrt{S_V}$ at the output and the device
responsivity $\partial V_{out}/\partial P_{heater}$. Another is to measure the
applied power at the input $\delta P_{in}$, and the $SNR$ at the output as
suggested by rearranging the above expression

\[  NEP  \equiv \frac{\sqrt{S_V}} {\delta V_{out}} * \delta P_{in} = \frac{\delta P_{in}}{ SNR} = \frac{\delta P_{heater}}{ 4 \times SNR} \]

In the above equation, the $SNR$ is in units of $\frac{V}{V/\sqrt{Hz}}$ and
$P_{in}=P_{heater}/4$. The latter expression is true since we have implicitly
assumed $P_{in}$ is that input power that produces the measurable $V_{out}$
signal. In our case, only one quarter of the heat power $P_{heater}$ injected at
the heater port produces the measured sideband signal.

\subsection{Experimental Design}

To measure the $NEP$, we use the measurement setup in Fig.~\ref{exfig: nep}(a,b)
and perform the following procedure:

\begin{itemize}
  \item We apply a carrier tone on the microwave line (Fig.~\ref{exfig: nep}(a),
        first panel). The $S_{21}$ parameter is the transfer function which
        determines the magnitude and phase of the signal at the output. Thus, a
        carrier tone at the resonant frequency, i.e., at the maximal dip of the
        $S_{21}$ parameter will have a smaller transmitted magnitude than a
        carrier tone placed off-resonance.

  \item Measurement of
        $P_{heater}(\omega) = I_{heater}(\omega) \times V_{heater}(\omega)$ is
        achieved by sourcing a current $I_{heater}(\omega)$ to the heater port
        and measuring the voltage drop
        $V_{heater}(\omega) = I_{heater}(\omega) R_{heater}$ over the heater
        port in a 4-wire lock-in measurement. Since we apply an AC heater
        current $I_{heater}(\omega) \propto \cos{\omega t}$, it follows that
        $P_{heater}(\omega) \propto \cos^2{\omega t} = \frac{1}{2}(1 + \cos(2 \omega t))$.
        Only the $2 \omega$ term in the final expression contributes to the
        $V_{out}$ sideband signal.

  \item Applying an AC heat power $P_{heater}$ to the heater port modulates the
        $S_{21}$ parameter between unheated and heated states (Fig.~\ref{exfig:
        nep}(a), second panel). Consistent with the heating measurements
        performed in the main text, the heated state has a lower resonant
        frequency and lower quality factor than the unheated state. The
        $2 \omega$ component of the input power $P_{heater}$ modulates the flake
        temperature at $2 \omega$. Thus, modulation of the $S_{21}$ resonance
        feature will occur at $2 \omega$.

  \item Placing the frequency of the carrier tone within the bandwidth of the
        modulated $S_{21}$ resonance feature will amplitude modulate the
        carrier, producing sidebands spaced at $2 \omega$ from the carrier
        (Fig.~\ref{exfig: nep}(a), second and third panel). Provided that the
        device is in the linear-response regime, the voltage of the sidebands
        will increase in proportion to applied heat power, i.e.
        $V_{sb} \propto P_{heater}$. It follows that the power of the sidebands
        will increase as $P_{sb} \propto P_{heater}^2$.

  \item The amplitude-modulated carrier is read out by a spectrum analyzer
        (Fig.~\ref{exfig: nep}(a), third panel). The signal-to-noise ratio of
        the sideband is used to calculate the $NEP$. We note that only one
        sideband is used in the $NEP$ measurement.
\end{itemize}

\subsection{Sideband Spectrum}
In Fig.~\ref{exfig: nep}(c), we see that application of an AC heater current of
magnitude $I_{heater}=20$ nA results in sidebands at 2$\omega$ offset from the
carrier, where $\omega =2 \pi \times 337 \, \rm{ Hz}$. In addition to the
2$\omega$ sidebands, sidebands at multiples of the $60 \, \rm{Hz}$ line
frequency frequency are present. Additionally, there are sidebands at $\omega$
approximately $10 \, \rm{dB}$ down from the 2$\omega$ sidebands. This can be
explained by a small DC offset in the heater current.

With increasing heater power, the magnitude of the sidebands saturates at a
value consistent with expectations. It is straightforward to show that a
resonance dip of $3 \, \rm{dB}$ generates a maximum amplitude modulation index
$m = 17\%$, which should produce sidebands $21 \, \rm{dB}$ lower than the
carrier. This is in agreement with the measured sideband magnitude that is
$23 \, \rm{dB}$ lower than the carrier.

\subsection{Sideband Scaling}

In the linear response regime, $V_{sb}\propto \delta P_{heater}$. Therefore, the
sideband signal as measured on the spectrum analyzer (in power units) should
scale as $P_{sb} \propto P^2_{heater}$, or by $20 \, \rm{dB/decade}$. This is
seen in Fig.~\ref{exfig: nep}(d) for applied heat $P_{heater}$ in the range
$-120 \, \rm{dBm}$ to $-105 \, \rm{dBm}$, where the slope of fit at
low-$P_{heater}$ is consistent with a scaling exponent $n=2$. This confirms that
our measurement is in linear-response regime at low $P_{heater}$. For greater
applied $P_{heater}$, the sideband power saturates as the amplitude modulation
reaches the full maximum of the resonance dip.

\subsection{$\mathbf{NEP}$ vs. $\mathbf{P_{heater}}$}

In the linear response regime, the $NEP$ is constant with respect to
$P_{heater}$ since $V_{sb} \propto P_{heater}$. This is shown Fig.~\ref{exfig:
  nep}e for $P_{heater} < -105 \, \rm{dBm}$. As stated above, the $NEP$ rises
for $P_{heater} > -105 \, \rm{dBm}$ as the $SNR$ saturates while $P_{heater}$
continues to increase.

\subsection{$\mathbf{NEP}$ vs. Carrier Frequency and Carrier Power}
To explore the $NEP$ as a function of the carrier tone, we generate a heat map
with swept carrier frequency $f_c$ and carrier power $P_c$ (Fig.~\ref{exfig:
  nep}(f,g)). For the lowest carrier powers, the $NEP$ is minimized for carrier
frequencies close to the resonance minimum, where the responsivity of the
resonance to applied heater power is greatest and therefore the amplitude
modulation of the carrier is greatest. As the carrier power $P_c$ is increased,
the junction is driven to nonlinearity, resulting in a resonance dip with a
steep falling edge and a shallow rising edge. This has the effect of enhancing
the $NEP$ on the falling edge and reducing it on the rising edge. For carrier
powers $P_c > -98 \, \rm{ dBm}$, the quality factor of the resonance feature is
degraded to such an extent that the amplitude modulation of sideband is reduced
and the $NEP$ increases. The $NEP$ reaches a minimum value of
$7\times10^{-17} \, \rm{W/}\sqrt{\, \rm{Hz}} $ for a carrier power
$P_{carrier}= -102$ dBm and carrier frequency $ f_{carrier}= 753.5$ MHz.

\subsection{Detection Limits}

The measured minimum noise-equivalent power
$NEP_{min} \approx 7\times10^{-17} \, \rm{W/}\sqrt{\rm{Hz}}$. It is limited by
the noise of the 4K cryoamp and is $\sim20\times$ larger than the thermal
fluctuation-limited $NEP= \sqrt{4k_BT^2G_{th}}$ at $T_{mxc} = 200 \, \rm{mK}$.
At $T_{mxc}=58 \, \rm{mK}$ , the projected thermal fluctuation-limited
$NEP_{proj} \approx 1\times10^{-19} \, \rm{W/\sqrt{Hz}}$, assuming that the
$T^4$ dependence of $G_{th}$ holds down to these temperatures
\cite{matherBolometerNoiseNonequilibrium1982,moseleyThermalDetectorsRay1984}.
The corresponding thermal fluctuation-limited energy resolution
$\delta E = NEP_{proj}\sqrt{\tau_{th}} \approx h \times 65 \, \rm{GHz}$,
assuming the projected thermal time constant
$\tau_{th} = \frac{C_{th}}{G_{th}} \approx 170 \, \rm{ns}$,
$n_{carrier}= \frac{10^{12}}{cm^2}$, $A= 25 \, \mu m^2$.


\begin{thebibliography}{10}
\expandafter\ifx\csname url\endcsname\relax
  \def\url#1{\texttt{#1}}\fi
\expandafter\ifx\csname urlprefix\endcsname\relax\def\urlprefix{URL }\fi
\providecommand{\bibinfo}[2]{#2}
\providecommand{\eprint}[2][]{\url{#2}}

\bibitem{fongUltrasensitiveWideBandwidthThermal2012}
\bibinfo{author}{Fong, K.~C.} \& \bibinfo{author}{Schwab, K.~C.}
\newblock \bibinfo{title}{Ultrasensitive and {{Wide-Bandwidth Thermal
  Measurements}} of {{Graphene}} at {{Low Temperatures}}}.
\newblock \emph{\bibinfo{journal}{Physical Review X}}
  \textbf{\bibinfo{volume}{2}}, \bibinfo{pages}{031006} (\bibinfo{year}{2012}).

\bibitem{fongMeasurementElectronicThermal2013}
\bibinfo{author}{Fong, K.~C.} \emph{et~al.}
\newblock \bibinfo{title}{Measurement of the {{Electronic Thermal Conductance
  Channels}} and {{Heat Capacity}} of {{Graphene}} at {{Low Temperature}}}.
\newblock \emph{\bibinfo{journal}{Physical Review X}}
  \textbf{\bibinfo{volume}{3}}, \bibinfo{pages}{041008} (\bibinfo{year}{2013}).

\bibitem{borzenetsBallisticGrapheneJosephson2016}
\bibinfo{author}{Borzenets, I.~V.} \emph{et~al.}
\newblock \bibinfo{title}{Ballistic {{Graphene Josephson Junctions}} from the
  {{Short}} to the {{Long Junction Regimes}}}.
\newblock \emph{\bibinfo{journal}{Physical Review Letters}}
  \textbf{\bibinfo{volume}{117}}, \bibinfo{pages}{237002}
  (\bibinfo{year}{2016}).

\bibitem{caladoBallisticJosephsonJunctions2015}
\bibinfo{author}{Calado, V.~E.} \emph{et~al.}
\newblock \bibinfo{title}{Ballistic {{Josephson}} junctions in edge-contacted
  graphene}.
\newblock \emph{\bibinfo{journal}{Nature Nanotechnology}}
  \textbf{\bibinfo{volume}{10}}, \bibinfo{pages}{761--764}
  (\bibinfo{year}{2015}).

\bibitem{draelosSupercurrentFlowMultiterminal2019}
\bibinfo{author}{Draelos, A.~W.} \emph{et~al.}
\newblock \bibinfo{title}{Supercurrent {{Flow}} in {{Multiterminal Graphene
  Josephson Junctions}}}.
\newblock \emph{\bibinfo{journal}{Nano Letters}} \textbf{\bibinfo{volume}{19}},
  \bibinfo{pages}{1039--1043} (\bibinfo{year}{2019}).

\bibitem{schmidtBallisticGrapheneSuperconducting2018}
\bibinfo{author}{Schmidt, F.~E.}, \bibinfo{author}{Jenkins, M.~D.},
  \bibinfo{author}{Watanabe, K.}, \bibinfo{author}{Taniguchi, T.} \&
  \bibinfo{author}{Steele, G.~A.}
\newblock \bibinfo{title}{A ballistic graphene superconducting microwave
  circuit}.
\newblock \emph{\bibinfo{journal}{Nature Communications}}
  \textbf{\bibinfo{volume}{9}}, \bibinfo{pages}{4069} (\bibinfo{year}{2018}).

\bibitem{borzenetsPhononBottleneckGrapheneBased2013}
\bibinfo{author}{Borzenets, I.~V.} \emph{et~al.}
\newblock \bibinfo{title}{Phonon {{Bottleneck}} in {{Graphene-Based Josephson
  Junctions}} at {{Millikelvin Temperatures}}}.
\newblock \emph{\bibinfo{journal}{Physical Review Letters}}
  \textbf{\bibinfo{volume}{111}}, \bibinfo{pages}{027001}
  (\bibinfo{year}{2013}).

\bibitem{leeGraphenebasedJosephsonJunction2020}
\bibinfo{author}{Lee, G.-H.} \emph{et~al.}
\newblock \bibinfo{title}{Graphene-based {{Josephson}} junction microwave
  bolometer}.
\newblock \emph{\bibinfo{journal}{Nature}} \textbf{\bibinfo{volume}{586}},
  \bibinfo{pages}{42--46} (\bibinfo{year}{2020}).

\bibitem{halbertalImagingResonantDissipation2017}
\bibinfo{author}{Halbertal, D.} \emph{et~al.}
\newblock \bibinfo{title}{Imaging resonant dissipation from individual atomic
  defects in graphene}.
\newblock \emph{\bibinfo{journal}{Science}} \textbf{\bibinfo{volume}{358}},
  \bibinfo{pages}{1303--1306} (\bibinfo{year}{2017}).

\bibitem{sairaDispersiveThermometryJosephson2016}
\bibinfo{author}{Saira, O.-P.}, \bibinfo{author}{Zgirski, M.},
  \bibinfo{author}{Viisanen, K.~L.}, \bibinfo{author}{Golubev, D.~S.} \&
  \bibinfo{author}{Pekola, J.~P.}
\newblock \bibinfo{title}{Dispersive {{Thermometry}} with a {{Josephson
  Junction Coupled}} to a {{Resonator}}}.
\newblock \emph{\bibinfo{journal}{Physical Review Applied}}
  \textbf{\bibinfo{volume}{6}}, \bibinfo{pages}{024005} (\bibinfo{year}{2016}).

\bibitem{kongResonantElectronlatticeCooling2018}
\bibinfo{author}{Kong, J.~F.}, \bibinfo{author}{Levitov, L.},
  \bibinfo{author}{Halbertal, D.} \& \bibinfo{author}{Zeldov, E.}
\newblock \bibinfo{title}{Resonant electron-lattice cooling in graphene}.
\newblock \emph{\bibinfo{journal}{Physical Review B}}
  \textbf{\bibinfo{volume}{97}}, \bibinfo{pages}{245416}
  (\bibinfo{year}{2018}).

\bibitem{tikhonovResonantSupercollisionsElectronphonon2018}
\bibinfo{author}{Tikhonov, K.~S.}, \bibinfo{author}{Gornyi, I.~V.},
  \bibinfo{author}{Kachorovskii, V.~Y.} \& \bibinfo{author}{Mirlin, A.~D.}
\newblock \bibinfo{title}{Resonant supercollisions and electron-phonon heat
  transfer in graphene}.
\newblock \emph{\bibinfo{journal}{Physical Review B}}
  \textbf{\bibinfo{volume}{97}}, \bibinfo{pages}{085415}
  (\bibinfo{year}{2018}).

\bibitem{wangCoherentControlHybrid2019}
\bibinfo{author}{Wang, J. I.-J.} \emph{et~al.}
\newblock \bibinfo{title}{Coherent control of a hybrid superconducting circuit
  made with graphene-based van der {{Waals}} heterostructures}.
\newblock \emph{\bibinfo{journal}{Nature Nanotechnology}}
  \textbf{\bibinfo{volume}{14}}, \bibinfo{pages}{120--125}
  (\bibinfo{year}{2019}).

\bibitem{tinkhamIntroductionSuperconductivity2004}
\bibinfo{author}{Tinkham, M.}
\newblock \emph{\bibinfo{title}{Introduction to Superconductivity}}
  (\bibinfo{publisher}{{Publisher: Dover Publications}}, \bibinfo{year}{2004}).

\bibitem{viljasElectronphononHeatTransfer2010}
\bibinfo{author}{Viljas, J.~K.} \& \bibinfo{author}{Heikkil{\"a}, T.~T.}
\newblock \bibinfo{title}{Electron-phonon heat transfer in monolayer and
  bilayer graphene}.
\newblock \emph{\bibinfo{journal}{Physical Review B}}
  \textbf{\bibinfo{volume}{81}}, \bibinfo{pages}{245404}
  (\bibinfo{year}{2010}).

\bibitem{chenElectronphononMediatedHeat2012}
\bibinfo{author}{Chen, W.} \& \bibinfo{author}{Clerk, A.~A.}
\newblock \bibinfo{title}{Electron-phonon mediated heat flow in disordered
  graphene}.
\newblock \emph{\bibinfo{journal}{Physical Review B}}
  \textbf{\bibinfo{volume}{86}}, \bibinfo{pages}{125443}
  (\bibinfo{year}{2012}).

\bibitem{roukesHotElectronsEnergy1985}
\bibinfo{author}{Roukes, M.~L.}, \bibinfo{author}{Freeman, M.~R.},
  \bibinfo{author}{Germain, R.~S.}, \bibinfo{author}{Richardson, R.~C.} \&
  \bibinfo{author}{Ketchen, M.~B.}
\newblock \bibinfo{title}{Hot electrons and energy transport in metals at
  millikelvin temperatures}.
\newblock \emph{\bibinfo{journal}{Physical Review Letters}}
  \textbf{\bibinfo{volume}{55}}, \bibinfo{pages}{422--425}
  (\bibinfo{year}{1985}).

\bibitem{halbertalNanoscaleThermalImaging2016}
\bibinfo{author}{Halbertal, D.} \emph{et~al.}
\newblock \bibinfo{title}{Nanoscale thermal imaging of dissipation in quantum
  systems}.
\newblock \emph{\bibinfo{journal}{Nature}} \textbf{\bibinfo{volume}{539}},
  \bibinfo{pages}{407--410} (\bibinfo{year}{2016}).

\bibitem{kokkoniemiBolometerOperatingThreshold2020}
\bibinfo{author}{Kokkoniemi, R.} \emph{et~al.}
\newblock \bibinfo{title}{Bolometer operating at the threshold for circuit
  quantum electrodynamics}.
\newblock \emph{\bibinfo{journal}{Nature}} \textbf{\bibinfo{volume}{586}},
  \bibinfo{pages}{47--51} (\bibinfo{year}{2020}).

\bibitem{dayBroadbandSuperconductingDetector2003}
\bibinfo{author}{Day, P.~K.}, \bibinfo{author}{LeDuc, H.~G.},
  \bibinfo{author}{Mazin, B.~A.}, \bibinfo{author}{Vayonakis, A.} \&
  \bibinfo{author}{Zmuidzinas, J.}
\newblock \bibinfo{title}{A broadband superconducting detector suitable for use
  in large arrays}.
\newblock \emph{\bibinfo{journal}{Nature}} \textbf{\bibinfo{volume}{425}},
  \bibinfo{pages}{817--821} (\bibinfo{year}{2003}).

\bibitem{wanduiThermalKineticInductance2020}
\bibinfo{author}{Wandui, A.} \emph{et~al.}
\newblock \bibinfo{title}{Thermal kinetic inductance detectors for
  millimeter-wave detection}.
\newblock \emph{\bibinfo{journal}{Journal of Applied Physics}}
  \textbf{\bibinfo{volume}{128}}, \bibinfo{pages}{044508}
  (\bibinfo{year}{2020}).

\bibitem{walshGrapheneBasedJosephsonJunctionSinglePhoton2017}
\bibinfo{author}{Walsh, E.~D.} \emph{et~al.}
\newblock \bibinfo{title}{Graphene-{{Based Josephson-Junction Single-Photon
  Detector}}}.
\newblock \emph{\bibinfo{journal}{Physical Review Applied}}
  \textbf{\bibinfo{volume}{8}}, \bibinfo{pages}{024022} (\bibinfo{year}{2017}).

\bibitem{lara-avilaQuantumlimitedCoherentDetection2019}
\bibinfo{author}{{Lara-Avila}, S.} \emph{et~al.}
\newblock \bibinfo{title}{Towards quantum-limited coherent detection of
  terahertz waves in charge-neutral graphene}.
\newblock \emph{\bibinfo{journal}{Nature Astronomy}}
  \textbf{\bibinfo{volume}{3}}, \bibinfo{pages}{983--988}
  (\bibinfo{year}{2019}).

\bibitem{hochbergDirectionalDetectionDark2017}
\bibinfo{author}{Hochberg, Y.}, \bibinfo{author}{Kahn, Y.},
  \bibinfo{author}{Lisanti, M.}, \bibinfo{author}{Tully, C.~G.} \&
  \bibinfo{author}{Zurek, K.~M.}
\newblock \bibinfo{title}{Directional detection of dark matter with
  two-dimensional targets}.
\newblock \emph{\bibinfo{journal}{Physics Letters B}}
  \textbf{\bibinfo{volume}{772}}, \bibinfo{pages}{239--246}
  (\bibinfo{year}{2017}).

\bibitem{kimDetectingKeVRangeSuperLight2020}
\bibinfo{author}{Kim, D.}, \bibinfo{author}{Park, J.-C.},
  \bibinfo{author}{Fong, K.~C.} \& \bibinfo{author}{Lee, G.-H.}
\newblock \bibinfo{title}{Detecting {{keV-Range Super-Light Dark Matter Using
  Graphene Josephson Junction}}}.
\newblock \emph{\bibinfo{journal}{arXiv:2002.07821 [cond-mat, physics:hep-ex,
  physics:hep-ph]}}  (\bibinfo{year}{2020}).
\newblock \eprint{2002.07821}.

\bibitem{mcallisterORGANExperimentAxion2017}
\bibinfo{author}{McAllister, B.~T.} \emph{et~al.}
\newblock \bibinfo{title}{The {{ORGAN}} experiment: {{An}} axion haloscope
  above 15 {{GHz}}}.
\newblock \emph{\bibinfo{journal}{Physics of the Dark Universe}}
  \textbf{\bibinfo{volume}{18}}, \bibinfo{pages}{67--72}
  (\bibinfo{year}{2017}).

\bibitem{baracchiniPTOLEMYProposalThermal2018}
\bibinfo{author}{Baracchini, E.} \emph{et~al.}
\newblock \bibinfo{title}{{{PTOLEMY}}: {{A Proposal}} for {{Thermal Relic
  Detection}} of {{Massive Neutrinos}} and {{Directional Detection}} of {{MeV
  Dark Matter}}}.
\newblock \emph{\bibinfo{journal}{arXiv:1808.01892 [astro-ph, physics:hep-ex,
  physics:physics]}}  (\bibinfo{year}{2018}).
\newblock \eprint{1808.01892}.

\bibitem{roukesYoctocalorimetryPhononCounting1999}
\bibinfo{author}{Roukes, M.~L.}
\newblock \bibinfo{title}{Yoctocalorimetry: Phonon counting in nanostructures}.
\newblock \emph{\bibinfo{journal}{Physica B: Condensed Matter}}
  \textbf{\bibinfo{volume}{263--264}}, \bibinfo{pages}{1--15}
  (\bibinfo{year}{1999}).

\bibitem{matherBolometerNoiseNonequilibrium1982}
\bibinfo{author}{Mather, J.~C.}
\newblock \bibinfo{title}{Bolometer noise: Nonequilibrium theory}.
\newblock \emph{\bibinfo{journal}{Applied Optics}}
  \textbf{\bibinfo{volume}{21}}, \bibinfo{pages}{1125--1129}
  (\bibinfo{year}{1982}).

\bibitem{geerlingsImprovingQualityFactor2012}
\bibinfo{author}{Geerlings, K.} \emph{et~al.}
\newblock \bibinfo{title}{Improving the quality factor of microwave compact
  resonators by optimizing their geometrical parameters}.
\newblock \emph{\bibinfo{journal}{Applied Physics Letters}}
  \textbf{\bibinfo{volume}{100}}, \bibinfo{pages}{192601}
  (\bibinfo{year}{2012}).

\bibitem{m.pozarMicrowaveEngineering4th2011}
\bibinfo{author}{M.~Pozar, D.}
\newblock \emph{\bibinfo{title}{Microwave {{Engineering}}, 4th {{Edition}}}}
  (\bibinfo{publisher}{Wiley}, \bibinfo{year}{2011}).

\bibitem{vanduzerPrinciplesSuperconductiveDevices1998}
\bibinfo{author}{Van~Duzer, T.} \& \bibinfo{author}{W~Turner, C.}
\newblock \emph{\bibinfo{title}{Principles of {{Superconductive Devices}} and
  {{Circuits}}}} (\bibinfo{publisher}{{Pearson}}, \bibinfo{year}{1998}).

\bibitem{leeUltimatelyShortBallistic2015}
\bibinfo{author}{Lee, G.-H.}, \bibinfo{author}{Kim, S.}, \bibinfo{author}{Jhi,
  S.-H.} \& \bibinfo{author}{Lee, H.-J.}
\newblock \bibinfo{title}{Ultimately short ballistic vertical graphene
  {{Josephson}} junctions}.
\newblock \emph{\bibinfo{journal}{Nature Communications}}
  \textbf{\bibinfo{volume}{6}}, \bibinfo{pages}{6181} (\bibinfo{year}{2015}).

\bibitem{nandaCurrentPhaseRelationBallistic2017}
\bibinfo{author}{Nanda, G.} \emph{et~al.}
\newblock \bibinfo{title}{Current-{{Phase Relation}} of {{Ballistic Graphene
  Josephson Junctions}}}.
\newblock \emph{\bibinfo{journal}{Nano Letters}} \textbf{\bibinfo{volume}{17}},
  \bibinfo{pages}{3396--3401} (\bibinfo{year}{2017}).

\bibitem{voutilainenEnergyRelaxationGraphene2011}
\bibinfo{author}{Voutilainen, J.} \emph{et~al.}
\newblock \bibinfo{title}{Energy relaxation in graphene and its measurement
  with supercurrent}.
\newblock \emph{\bibinfo{journal}{Physical Review B}}
  \textbf{\bibinfo{volume}{84}}, \bibinfo{pages}{045419}
  (\bibinfo{year}{2011}).

\bibitem{giazottoOpportunitiesMesoscopicsThermometry2006}
\bibinfo{author}{Giazotto, F.}, \bibinfo{author}{Heikkil{\"a}, T.~T.},
  \bibinfo{author}{Luukanen, A.}, \bibinfo{author}{Savin, A.~M.} \&
  \bibinfo{author}{Pekola, J.~P.}
\newblock \bibinfo{title}{Opportunities for mesoscopics in thermometry and
  refrigeration: {{Physics}} and applications}.
\newblock \emph{\bibinfo{journal}{Reviews of Modern Physics}}
  \textbf{\bibinfo{volume}{78}}, \bibinfo{pages}{217--274}
  (\bibinfo{year}{2006}).

\bibitem{timofeevRecombinationLimitedEnergyRelaxation2009}
\bibinfo{author}{Timofeev, A.~V.} \emph{et~al.}
\newblock \bibinfo{title}{Recombination-{{Limited Energy Relaxation}} in a
  {{Bardeen-Cooper-Schrieffer Superconductor}}}.
\newblock \emph{\bibinfo{journal}{Physical Review Letters}}
  \textbf{\bibinfo{volume}{102}}, \bibinfo{pages}{017003}
  (\bibinfo{year}{2009}).

\bibitem{moseleyThermalDetectorsRay1984}
\bibinfo{author}{Moseley, S.~H.}, \bibinfo{author}{Mather, J.~C.} \&
  \bibinfo{author}{McCammon, D.}
\newblock \bibinfo{title}{Thermal detectors as x-ray spectrometers}.
\newblock \emph{\bibinfo{journal}{Journal of Applied Physics}}
  \textbf{\bibinfo{volume}{56}}, \bibinfo{pages}{1257--1262}
  (\bibinfo{year}{1984}).

\end{thebibliography}
\end{document}